\documentclass{article}

\usepackage[english]{babel}

\newcommand\np{\mbox{NP}}

\newtheorem{theorem}{Theorem}{\bf}{\it }
\newtheorem{conjecture}{Conjecture}{\bf}{\it }
\newtheorem{lemma}{Lemma}{\bf}{\it }
\newtheorem{corollary}{Corollary}{\bf}{\it }
{\bf}{\it }
{\bf}{\it }

\title{Experimental Algorithm for the Maximum Independent Set Problem}

\author{Anatoly D. Plotnikov\footnote{
E-mail address: {\tt akplokt(at)kvnk.ua (remove k's)}}}

\date{}

\begin{document}
\maketitle

\begin{abstract}
We develop an experimental algorithm for the exact solving of the maximum independent 
set problem. The algorithm consecutively finds the maximal independent sets of 
vertices in an arbitrary undirected graph such that the next such set contains more 
elements than the preceding one. For this purpose, we use a technique, developed 
by Ford and Fulkerson for the finite partially ordered sets, in particular, 
their method for partition of a poset into the minimum number of chains with 
finding the maximum antichain. In the process of solving, a special digraph is
constructed, and a conjecture is formulated concerning properties of such digraph.
This allows to offer of the solution algorithm. Its theoretical estimation of
running time equals to is $O(n^{8})$, where $n$ is the number of
graph vertices. The offered algorithm was tested by a
program on random graphs. The testing the confirms correctness of the algorithm.
\end{abstract}

\vspace{1pc}

\np-complete, the class \np, a polynomial-time algorithm, 
the partially ordered set.

\maketitle

\section{Introduction}

A set of all undirected $n$-vertex graphs without loops and multiple edges
is denoted by $L_{n}$.

Let there be a graph $G=(V,\Gamma)\in L_{n}$, where $V$ is the set of graph 
vertices, and $\Gamma$ is a mapping from $V$ to $V$. Any subgraph $Q_{1}$ =
$(V_{1},\Gamma_{1})$ of $G$ is called a {\em clique} induced by $V_{1}$, if
\[
\Gamma_{1}(v)=V_{1}\setminus \{v\}
\]
for all $v\in V_{1}$. In a special case, when $Card(V_{1}) = 1$, the 
one-vertex subgraph $Q_{1}=(V_{1},\Gamma_{1})$ is called a 
{\it single-vertex clique}.

A clique $Q_{1}$ is called {\it maximal} if any vertex $v\in V$ cannot be
attached to it so that the new vertex set also has formed a clique of the 
graph $G$. A clique ${\hat Q}$ is called {\it maximum} if the graph $G$ 
has not a clique of the greater size than ${\hat Q}$.

It be required to find the maximum clique of a graph $G\in L_{n}$.
The known problem of finding the maximum clique has formulated \cite{gary-johnson}.

A vertex set $U\subseteq V$ of $G$ is called {\it independent} if
\[
U\cap \Gamma (U)=\oslash.
\]
An independent set $U$ of graph vertices is called {\it maximal (MIS)} if
\[
U\cup \Gamma (U)=V.
\]

A MIS $\tilde U$ is called {\it maximum (MMIS)} if $Card(\tilde U)\geq Card(U)$
for any MIS $U$ of $G$.

It be required to find the MMIS of a graph $G$. Again, we have formulated
well-know the maximum independent set problem (MISP). 

Both of the formulated above problems are \np-complete \cite {gary-johnson, 
compendium}. They are closely connected with each other: the solution 
of one of them attracts the solution another.

A graph ${\bar G}$ is called {\it complement} to the graph $G$ if it
has the same vertex set, and edges join two vertices of the graph 
${\bar G}$ iff these vertices are non-adjacent in $G$.

It is not difficult to see that any clique of $G$ corresponds to 
the independent set of graph vertices in ${\bar G}$, and conversely.
Therefore, the finding the maximum independent set in one graph is
equivalent to the finding the maximum clique in its complement graph.
In the given work, we examine the maximum independent set problem.

The intention of the given paper is to design a polynomial-time 
algorithm for the exact solving of the maximum independent set problem in
an arbitrary undirected graph. For this purpose, we use a technique, 
developed by Ford and Fulkerson (see \cite{ford-fulkerson} and Appendix A) 
for the finite 
partially ordered sets, in particular, their method of partition poset into 
the minimum number of chains with finding the maximum antichain. In the 
process of the solving, a special digraph is constructed, and a conjecture is 
formulated concerning properties such digraph. This allows to offer a solution 
algorithm, a theoretical estimation of running time for which equals to is
$O(n^{8})$, where $n$ is the number of graph vertices. The offered algorithm 
was tested by a program. The testing confirms the correctness 
of the algorithm.

Notice that the author did not aspire to creation of the most optimum and fast 
algorithm. In work \cite{tarjan}, Tarjan has presented a table, which 
shows a dynamics of perfecting a complexity evaluation for solving 
some problems. Whence one can make a conclusion that after the appearance 
of an initial algorithm for solving certain problem, its improvement is found
quickly. It is presented that proposed algorithm also can be improved 
hereinafter.

\section{The Basic Definitions}
\label{init}

Let there is a graph $G=(V,\Gamma)\in L_{n}$. We will partition the vertex 
set $V$ into subsets
\begin{equation}
\label{inpart}
V^{0}, V^{1}, \ldots , V^{m}
\end{equation}
in such a way that a subset $V^{k}$ ($k=0, 1,\ldots,m$) is a MIS of
a subgraph $G_{k}$ = ($V\setminus (V^{0}\cup V^{1}\cup \cdots \cup V^{k-1})$, 
$\Gamma_{k}$) = ($V^{k}\cup \cdots \cup V^{m}$, $\Gamma_{k}$). Clearly,
$G_{0}$ = ($V^{0}\cup \cdots \cup V^{m}$, $\Gamma_{0}$) = $G$.

By the given undirected graph $G$ and the partition (\ref{inpart}) we can 
construct a digraph ${\vec G}(V^{0})$ = ($V, {\vec \Gamma}$) in the following
way. If an edge of $G$ joins a vertex $v_{i}\in V^{k_{1}}$ with a vertex
$v_{j}\in V^{k_{2}}$ then this edge is replaced by an arc ($v_{i},v_{j}$) 
when $k_{1}<k_{2}$. The vertex $v_{i}$ is called the {\it tail} of
($v_{i},v_{j}$), and the vertex $v_{j}$ is called the {\it head} of this arc.

As the result we have an acyclic digraph ${\vec G}(V^{0})$ = 
($V,{\vec \Gamma}$). The set $V^{0}$ is called {\it initiating}.

In general case we can construct a set $D(G)$ of different acyclic digraphs
as it was indicated above. Each digraph of $D(G)$ corresponds to
the graph $G\in L_{n}$. Further we will consider {\it only} digraphs of $D(G)$.

The maximum length $\rho(v)$ of a directed path, connecting a vertex $v\in V$ 
with some vertex of the initiating set $V^{0}$, is called the {\it rank} of $v$. 
The set of all graph vertices having the same rank $\rho(v)=k$ is called the 
{\it $k$-th layer} of ${\vec G}(V^{0})$ and designated as $V^{k}$.

To apply the partially ordered set technique, each digraph ${\vec G}(V^{0})$
is assigned to a transitive closure graph (TCG) ${\vec G}_{t}(V^{0})$ = 
($V,{\vec \Gamma}_{t}$) \cite{swamy-thulasiraman, west}. As the digraph 
${\vec G}(V^{0})$ is acyclic and loopless, its transitive closure 
${\vec G}_{t}(V^{0})$ is a graph of a strict partial order ($V, >$).
Further, we will not distinguish the transitive closure graph 
${\vec G}_{t}(V^{0})$ and partially ordered set (poset) ($V, >$). 
Therefore, we will consider, for example, antichains of the TCG 
${\vec G}_{t}(V^{0})$.

There exists an efficient algorithm to construct the TCG. Its running time is
equal to $O(n^{3})$ (see, for example, \cite{swamy-thulasiraman, reingold}). 

An arc ($v_{i},v_{j}$) of ${\vec G}_{t}(V^{0})$ will be called {\it essential}
if there exists the arc ($v_{i},v_{j}$) of the digraph ${\vec G}(V^{0})$.
Otherwise, the arc ($v_{i},v_{j}$) will be called {\it fictitious}. An
essential arc is also designated as $v_{i}> v_{j}$, and a fictitious arc is 
also designated as $v_{i}\gg v_{j}$. Obviously that any fictitious arc 
of ${\vec G}_{t}(V^{0})$ determines two independent vertices of the digraph 
${\vec G}(V^{0})$.

Let there is a poset $(A, \geq)$.

If $a\geq b$ or $b\geq a$, the elements $a$ and $b$ of $A$ are called 
{\it comparable}. If $a\not\geq b$ and $b\not\geq a$, such pair of elements 
is called {\it incomparable}.

If $A_{1}\subseteq A$ and each pair of elements of $A_{1}$ is comparable, 
we shall say that $A_{1}$ determines {\it a chain} of $(A, \geq)$. If 
$A_{1}\subseteq A$ and each pair of elements of $A_{1}$ is incomparable, 
we shall say that $A_{1}$ be {\it an antichain} of $(A, \geq)$. The antichain
$A_{1}$ is {\it the maximum} in $(A, \geq)$, if $Card(A_{1})\geq Card(A^{*})$ 
for any antichain $A^{*}\subseteq A$ in $(A, \geq)$.

We say that poset $(A, \geq)$ is partitioned into chains $A_{1}$, \ldots,
$A_{p}$, if each $A_{i}$ ($A_{i}\not= \oslash$, $i=\overline {1, p}$) be 
a chain,
\[
\bigcup_{i=1}^{p} A_{i}=A,
\]
and $A_{i}\cap A_{j}=\oslash$, when $i\not= j$ ($i, j\in \{1, \ldots , p\}$).

The partition of the poset $(A, \geq)$ into chains is called {\it minimum}, 
if it has the minimum number of elements $p$ in comparison with other 
partitions of $(A, \geq)$ into chains. Such partition also is called 
{\it minimum chain partition (MCP)} of poset $(A, \geq)$.

As ${\vec G}_{t}(V^{0})$ is a graph of strict partial order, we can find 
MCP ${\cal P}$ = $\{S_{1}, \ldots , S_{p}\}$. In common case, this partition 
is ambiguous.

\begin{figure}[hbp]
\centering
\mbox{\unitlength=1.00mm
\special{em:linewidth 0.4pt}
\linethickness{0.4pt}
\begin{picture}(56.00,39.07)
\put(5.00,10.00){\line(0,1){25.00}}
\put(5.00,35.00){\line(2,-3){17.00}}
\put(5.15,35.22){\line(2,-3){17.00}}
\put(5.30,35.37){\line(2,-3){17.00}}
\put(4.85,34.92){\line(2,-3){17.00}}
\put(4.70,34.77){\line(2,-3){17.00}}
\put(21.73,10.07){\line(0,1){25.00}}
\put(21.73,35.07){\line(-2,-3){17.00}}
\put(21.88,34.92){\line(-2,-3){17.00}}
\put(22.03,34.77){\line(-2,-3){17.00}}
\put(21.68,35.22){\line(-2,-3){17.00}}
\put(21.53,35.37){\line(-2,-3){17.00}}
\put(5.00,10.07){\line(1,4){4.53}}
\put(22.00,10.07){\line(-1,4){4.53}}
\put(5.00,35.07){\circle*{2.50}}
\put(22.00,35.07){\circle*{2.50}}
\put(10.00,28.07){\circle*{2.50}}
\put(17.00,28.07){\circle*{2.50}}
\put(5.00,10.07){\circle*{2.50}}
\put(22.00,10.07){\circle*{2.50}}
\put(9.00,7.07){\makebox(0,0)[cc]{$v_{1}$}}
\put(26.00,7.07){\makebox(0,0)[cc]{$v_{2}$}}
\put(10.50,32.07){\makebox(0,0)[cc]{$v_{3}$}}
\put(16.50,32.07){\makebox(0,0)[cc]{$v_{4}$}}
\put(6.00,39.07){\makebox(0,0)[cc]{$v_{5}$}}
\put(21.00,39.07){\makebox(0,0)[cc]{$v_{6}$}}
\put(35.00,10.07){\line(0,1){25.00}}
\put(35.00,35.07){\line(2,-3){17.00}}
\put(35.15,35.22){\line(2,-3){5.00}}
\put(35.30,35.37){\line(2,-3){5.00}}
\put(34.85,34.92){\line(2,-3){5.00}}
\put(34.70,34.77){\line(2,-3){5.00}}
\put(51.73,10.07){\line(0,1){25.00}}
\put(51.73,35.07){\line(-2,-3){17.00}}
\put(51.88,34.92){\line(-2,-3){5.00}}
\put(52.03,34.77){\line(-2,-3){5.00}}
\put(51.58,35.22){\line(-2,-3){5.00}}
\put(51.43,35.37){\line(-2,-3){5.00}}
\put(35.00,10.07){\line(1,4){4.53}}
\put(35.15,9.92){\line(1,4){4.53}}
\put(35.30,9.77){\line(1,4){4.53}}
\put(34.85,10.22){\line(1,4){4.53}}
\put(34.70,10.37){\line(1,4){4.53}}
\put(52.00,10.07){\line(-1,4){4.53}}
\put(52.15,10.22){\line(-1,4){4.53}}
\put(52.30,10.37){\line(-1,4){4.53}}
\put(51.85,9.92){\line(-1,4){4.53}}
\put(51.70,9.77){\line(-1,4){4.53}}
\put(35.00,35.07){\circle*{2.50}}
\put(52.00,35.07){\circle*{2.50}}
\put(40.00,28.07){\circle*{2.50}}
\put(47.00,28.07){\circle*{2.50}}
\put(35.00,10.07){\circle*{2.50}}
\put(52.00,10.07){\circle*{2.50}}
\put(39.00,7.07){\makebox(0,0)[cc]{$v_{1}$}}
\put(56.00,7.07){\makebox(0,0)[cc]{$v_{2}$}}
\put(40.50,32.07){\makebox(0,0)[cc]{$v_{3}$}}
\put(46.50,32.07){\makebox(0,0)[cc]{$v_{4}$}}
\put(36.00,39.07){\makebox(0,0)[cc]{$v_{5}$}}
\put(51.00,39.07){\makebox(0,0)[cc]{$v_{6}$}}
\put(13.07,2.00){\makebox(0,0)[cc]{(a)}}
\put(43.07,2.00){\makebox(0,0)[cc]{(b)}}
\end{picture}
}
\caption{Different MCPs of the transitive closure graph}
\label{f1-3}
\end{figure}
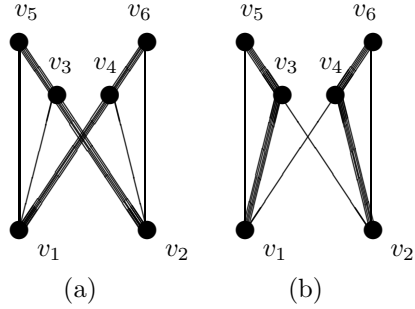

Different MCPs of the transitive closure graph is shown on the Fig. 
\ref{f1-3} (a) and (b). The digraph arcs, belonging chains of MCP, are 
represented by thick lines. Here and hereinafter, we suppose that 
orientation of arcs of the digraph is from below to upwards.

Let $V(S_{q})$ ($q=\overline {1, p}$) be the set of vertices, belonging to
the chain $S_{q}$. If vertices $v_{i},v_{j}\in V(S_{q})$ are endpoints for
a fictitious arc $v_{i}\gg v_{j}$ then the vertex $v_{j}$ is called 
{\it marked}. The set of all marked vertices of ${\vec G}_{t}(V^{0})$ is
determined by the found MCP $\cal P$ and it differs for different
MCPs. The set of all marked vertices of ${\vec G}_{t}(V^{0})$ is designated
as ${\cal B}({\cal P})$.

For example, for the MCP ${\cal P}_{1}$ represented in Fig. \ref{f1-3} (a), we 
have ${\cal B}({\cal P}_{1})$ = $\{v_{5},v_{6}\}$, and for the MCP 
${\cal P}_{2}$ represented in Fig. \ref{f1-3} (b), we have  
${\cal B}({\cal P}_{2})$ = $\oslash$.

\begin{lemma}
\label{mos}
Let ${\vec G}(V^{0})$ be a digraph, and ${\vec G}_{t}(V^{0})$ be its transitive
closure graph. If ${\cal B}({\cal P})=\oslash$, where ${\cal P}$ is an MCP of 
${\vec G}_{t}(V^{0})$, then the maximum antichain ${\cal U}$ is the MMIS of the
graph $G\in L_{n}$, and the MCP determines the minimum clique partition of $G$.
\end{lemma}

If conditions of Lemma \ref{mos} are satisfied then each chain 
$S_{q}\in {\cal P}$ is a clique of ${\vec G}(V^{0})$. Therefore, the MCP
${\cal P}$ is the minimum clique partition.

On the other hand, vertices of the maximum antichain ${\cal U}$ of 
${\vec G}_{t}(V^{0})$ belong to distinct cliques, that is, the number of
vertices in the MMIS is equal to the number of vertices in the set 
$\cal U$.\hfill Q.E.D.

\vspace{1pc}

Notice that Lemma \ref{mos} may be satisfied if the digraph ${\vec G}(V^{0})$ 
has no transitive orientation.

It is obvious that any antichain $U\subset V$ of ${\vec G}_{t}(V^{0})$ 
determines an independent vertex set of the digraph ${\vec G}(V^{0})$, and 
an independent vertex set $U\subset V$ of ${\vec G}(V^{0})$ determines an 
antichain of ${\vec G}_{t}(V^{0})$ if and only if no two vertices of $U$ 
belong to the same directed chain of ${\vec G}(V^{0})$.
\section{A vertex-saturated digraph}
\label{VS}

Let there is an acyclic digraph ${\vec G}(V^{0})$ = ($V, {\vec \Gamma}$).

Further, let $W\subset V$ be some independent vertex set. For the digraph 
${\vec G}(V^{0})$ we define an unary operation {\it cutting}
$\sigma_{W}({\vec G}(V^{0}))$. This operation consists of reorientation of
all arcs of ${\vec G}(V^{0})$ incoming into vertices of the set $W$.
It is easy to see that the result of this operation is also a digraph 
${\vec G}(Y^{0})$, where
\begin{displaymath}
Y^{0}=(V^{0}\setminus {\vec \Gamma}^{-1}(W))\cup W.
\end{displaymath}
Here, ${\vec \Gamma}^{-1}$ is a maping, inverse to ${\vec \Gamma}$.

\begin{theorem}
Let there be a digraph ${\vec G}(V^{0})\in D(G)$ and $W$ be some independent 
vertex set. Then a digraph ${\vec G}(Y^{0})$ = $\sigma_{W}({\vec G}(V^{0}))$
is also acyclic and ${\vec G}(Y^{0})\in D(G)$.
\end{theorem}

Indeed, since the digraph ${\vec G}(V^{0})$ = ($V, {\vec \Gamma}$) is acyclic 
then any its part is also an acyclic digraph. Thus, a directed subgraph
${\vec G}_{1}$ = ($V\setminus W, {\vec \Gamma}_{1}$) is acyclic.

Obviously, ${\vec G}_{1}\subset {\vec G}(Y^{0})$. Attach the independent 
vertex set $W$ to the subgraph ${\vec G}_{1}$. Join each vertex $x\in W$ 
with a vertex $y$ of ${\vec G}_{1}$ by the arc ($x, y$) if and only 
if there exists the arc ($y, x$) of the digraph ${\vec G}(V^{0})$. It is
evident that the resulting digraph ${\vec G}(Y^{0})$ is also acyclic.

At last, we have ${\vec G}(Y^{0})\in D(G)$ since any reorientation of 
arcs of the digraph ${\vec G}(V^{0})$ does not change independence relation of 
its vertices.\hfill Q.E.D.

\vspace{1pc}

Let there be a digraph ${\vec G}(V^{0})$ and its transitive closure graph 
${\vec G}_{t}(V^{0})$.

We can find the MCP ${\cal P}$ of the graph ${\vec G}_{t}(V^{0})$ constructing
the maximum antichain ${\cal U}$ simultaneously as it is described
in Appendix \ref{poset}.

In general case we can find some distinct maximum antichains.

We will say that an antichain ${U_{1}}$ {\it precedes} an antichain 
$U_{2}$ in the graph ${\vec G}_{t}(V^{0})$ and designate it as
$U_{1} \prec U_{2}$ if for all vertices $x \in U_{1} \setminus U_{2}$ 
there is a vertex $y \in U_{2} \setminus U_{1}$ such that $x \leq y$. 

By means of Ford and Fulkerson's methodology such a maximum
antichain ${\cal U}$ of the TCG ${\vec G}_{t}(V^{0})$ can be found
that precedes any other maximum antichain $U_{1}$, i.e. 
${\cal U}_{1} \prec {\cal U}$ for any antichain ${\cal U}_{1}$ of the TCG 
${\vec G}_{t}(V^{0})$. This antichain of the graph ${\vec G}_{t}(V^{0})$ 
we will call {\it general}.

In addition to the general antichain, we may find other maximum antichains of 
${\vec G}_{t}(V^{0})$ if they exist. So for any vertex $v \in V$ of the TCG 
${\vec G}_{t}(V^{0})$, it is possible to find a maximum antichain ${\cal U}(v)$ 
such that $v \in {\cal U}(v)$. Technically, to find the antichain 
${\cal U}(v)$, it is sufficient, in the adjacent matrix of 
${\vec G}_{t}(V^{0})$ containing the maximum number of units in allowable 
cells (and the marks are appointed by the Ford-Fulkerson's algorithm), 
to add the mark $(*)$ to the existing marks for a row, corresponding 
to vertex $v$, and to execute a cycle of appointment of marks. In this case,
in the first step of appointment of marks, all columns are marked, which
contain the admissible cells (incluging a chosen cell). Clearly, the antchain 
${\cal U}(v)$ will be general for the vertex $v$, that is, any other antichain, 
containing vertex $v$, will precede this antichain. 

\begin{figure}[h]
\centering
\mbox{\unitlength=1.00mm
\special{em:linewidth 0.4pt}
\linethickness{0.4pt}
\begin{picture}(72.00,34.00)
\put(33.00,10.00){\line(0,1){20.00}}
\put(33.00,30.00){\line(-2,-3){6.67}}
\put(26.33,20.00){\line(2,-3){6.67}}
\put(33.00,10.00){\line(1,2){10.00}}
\put(33.15,9.85){\line(1,2){10.00}}
\put(33.30,9.70){\line(1,2){10.00}}
\put(32.85,10.15){\line(1,2){10.00}}
\put(32.70,10.30){\line(1,2){10.00}}
\put(33.00,30.00){\line(1,-2){10.00}}
\put(33.15,30.15){\line(1,-2){10.00}}
\put(33.30,30.30){\line(1,-2){10.00}}
\put(32.85,29.85){\line(1,-2){10.00}}
\put(32.70,29.70){\line(1,-2){10.00}}
\put(33.00,30.00){\circle*{3.00}}
\put(43.00,30.00){\circle*{3.00}}
\put(26.00,20.00){\circle*{3.00}}
\put(33.00,10.00){\circle*{3.00}}
\put(43.00,10.00){\circle*{3.00}}
\put(30.00,7.00){\makebox(0,0)[cc]{$a_{1}$}}
\put(22.00,19.00){\makebox(0,0)[cc]{$a_{5}$}}
\put(30.00,34.00){\makebox(0,0)[cc]{$a_{2}$}}
\put(46.00,34.00){\makebox(0,0)[cc]{$a_{3}$}}
\put(46.00,7.00){\makebox(0,0)[cc]{$a_{4}$}}
\end{picture}
}
\caption{The partially ordered set}
\label{f1-2}
\end{figure}
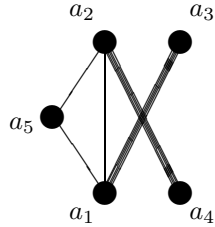

Notice that in general case it is not true that for any vertex $v \in V$ of 
the TCG ${\vec G}_{t}(V^{0})$ there exists a maximum antichain 
${\cal U}(v)$. For example, consider the digraph of the partially ordered set, 
shown in Fig. \ref{f1-2}. It is easy to see that there are no maximum
antichains ${\cal U}(a_{1})$ and ${\cal U}(a_{2})$ in it. 

A directed subgraph ${\vec G}(V^{k})$ $(k = \overline{0, m-1})$ of the 
digraph ${\vec G}(V^{0})$ we will call {\it saturated with respect 
to the initiating set $V^{k}$}, if, in its transitive closure graph
${\vec G}_{t}(V^{k})$, any maximum antichain ${\cal U}(v) \subset V$, when it 
exists, is a MIS of the subgraph ${\vec G}(V^{k})$ and satisfies the relation: 
$Card({\cal U}(v)) = Card(V^{k})$. Evidently, the subgraph 
${\vec G}(V^{m})$ has no arcs, and therefore it is saturated with
respect to its initiating set always.

A digraph ${\vec G}(V^{0})$ is called {\it vertex-saturated (VS-digraph)} if any 
of its directed subgraphs ${\vec G}(V^{k})$ $(k = \overline{0, m})$, induced 
by the layer $V^{k}$, is saturated with respect to the initiating set $V^{k}$.

Notice that digraph, represented in Fig. \ref{f1-3}, is not vertex-saturated.

Let there be some digraph ${\vec G}(V^{0})$. To construct a VS-digraph, we 
use the following algorithm.

\begin{list}{}{
\setlength{\topsep}{2mm}
\setlength{\itemsep}{0mm}
\setlength{\parsep}{1mm}
}
\item
{\bf The algorithm VS.}
\item[{\it Step 1.}] Put $k:=0$ and $\alpha :=$ {\bf false}.
\item[{\it Step 2.}] Find the transitive closure graph ${\vec G}_{t}(V^{k})$.
\item[{\it Step 3.}] Construct an MCP of ${\vec G}_{t}(V^{k})$.
\item[{\it Step 4.}] Find the maximum antichain of ${\vec G}_{t}(V^{k})$ for 
each vertex $v \in V^{k} \cup \cdots \cup V^m$.
\item[{\it Step 5.}] Check whether each of the found maximum antichains 
${\cal U}(v)$ of the graph ${\vec G}_{t}(V^{k})$ is a MIS of the digraph 
${\vec G}(V^{k})$ and $Card({\cal U}(v))=Card(V^{k})$. If it is true, finish 
the design of the digraph ${\vec G}(V^{k})$ saturated with respect to 
the initiating set $V^{k}$. Go to Step 6.

Otherwise complete the found antichain ${\cal U}(v)$ (when it is necessary) to 
a MIS, put $W:={\cal U}(v)$ and construct a new acyclic digraph ${\vec G}(V^{k})$ 
by the cutting operation $\sigma_{W}({\vec G}(V^{0}))$, put $\alpha:=$ {\bf true} 
and construct the new digraph ${\vec G}(V^{0})$. Return to Step 2.
\item[{\it Step 6.}] Compute $k:=k+1$. If $k<m$ then distinguish the transitive 
closure graph ${\vec G}_{t}(V^{k})$ from ${\vec G}_{t}(V^{k-1})$, keeping all 
chains of the MCP of ${\vec G}_{t}(V^{k-1})$ which are incident to vertices of 
the new graph. Go to Step 4.

If $k=m$ and $\alpha =$ {\bf true}, go to Step 1. If $k=m$ and $\alpha=$ 
{\bf false}, go to Step 7.
\item[{\it Step 7.}] Finish of this algorithm. A VS-digraph is constructed.
\end{list}

\vspace{1pc}

We will show that the algorithm VS constructs a vertex-saturated digraph.

\begin{theorem}
\label{tt}
Let ${\vec G}(V^{0})$ be a digraph constructed by the algorithm VS. Next,
let ${\vec G}(V^{k})$ = $(Y_{k}, \Gamma)$ $(k=\overline {0, m})$ be a
directed subgraph induced by the layer $V^{k}$. Then each antichain 
${\cal U} \subset Y_{k}\setminus V^{k}$ of the graph 
${\vec G}_{t}(V^{k})$ = $(Y_{k}, \Gamma_{t})$ obeys the following relation
\begin{equation}
\label{bp}
Card(V^{k}\cap \Gamma_{t}^{-1}{\cal U}) \geq Card(\cal U)
\end{equation}
\end{theorem}

Assume that, in the directed subgraph ${\vec G}(V^{k})$, an antichain 
${\cal U}\subset Y_{k}\setminus V^{k}$ of ${\vec G}_{t}(V^{k})$ = 
($Y_{k}, \Gamma_{t}$) will be found such that
\begin{displaymath}
Card(V^{k}\cap \Gamma_{t}^{-1}({\cal U})) < Card(\cal U).
\end{displaymath}

Construct a set 
${\cal U}^{*}$ = $(V^{k}\setminus \Gamma_{t}^{-1}(\cal U)) \cup {\cal U}$. 
Clearly, the set ${\cal U}^{*}$ is independent in the directed
subdigraph ${\vec G}(V^{k})$ and $Card({\cal U}^{*})> Card(V^{k})$.

Since the set $V^{k}$ is an antichain of the TCG ${\vec G}_{t}(V^{k})$ = 
($Y_{k}, \Gamma_{t}$) then the set 
$V^{k}\setminus \Gamma_{t}^{-1}({\cal U})\subset V^{k}$ is an antichain of
this graph. The set ${\cal U}\subset Y_{k}\setminus V^{k}$ is an antichain
of ${\vec G}_{t}(V^{k})$ by conditions of Theorem \ref{tt}.

Obviously,
${\cal U}\cap \Gamma_{t}(V^{k}\setminus \Gamma_{t}^{-1}(\cal U))=\oslash$,
therefore the set ${\cal U}^{*}$ is an antichain of ${\vec G}_{t}(V^{k})$.

We have obtained a contradiction since the directed subgraph ${\vec G}(V^{k})$ 
is constructed by the algorithm VS and the antichain ${\cal U}^{*}$ could be
discovered in Step 5. This proves the validity of Theorem \ref{tt}.\hfill Q.E.D.

\begin{corollary}
The digraph ${\vec G}(V^{0})$, constructed by the algorithm VS, is 
vertex-saturated.
\end{corollary}

\begin{theorem}
VS-digraph can be constructed in time $O(n^{5})$.
\end{theorem}

One completion of the steps 1 -- 6 requires $O(n_{k}^{3})$ time units,
where $n_{k}$ is the size of the vertex set of ${\vec G}(V^{k})$. 
Assuming that for each completion of these steps the maximum antichain 
increases at one vertex, we obtain the design time of a digraph 
${\vec G}(V^{k})$ vertex-saturated with respect to the initiating set, is equal 
to $O(n_{k}^{4})$.

Hence, the design time of a correctly constructed vertex-saturated
digraph ${\vec G}(V^{0})$ is equal to:
\begin{displaymath}
\sum_{\forall n_{k}} O(n_{k}^{4}) = O(n^{5}).
\end{displaymath}
\hfill Q.E.D.

\begin{theorem}
\label{essential}
Let a digraph ${\vec G}(V^{0})$ be vertex-saturated. Then there exists 
an MCP ${\cal P}$ of the graph ${\vec G}_{t}(V^{0})$ such that its chains 
contain only essential arcs.
\end{theorem}

Let a digraph ${\vec G}(V^{0})$ is constructed by the algorithm VS.
By Theorem \ref{tt}, each bipartite digraph $G(V^{k},V^{k+1})$ 
($k=\overline {0, m}$) of this digraph satisfies the Hall's theorem
and, hence, has a matching that saturates each vertex of the set 
$V^{k+1})$.\hfill Q.E.D.

\begin{corollary}
\label{start}
Let ${\vec G}(V^{0})$ be a VS-digraph. Then each chain of an MPP ${\cal P}$ of
${\vec G}_{t}(V^{0})$ is begun by some vertex $v$ of $V^{0}$.
\end{corollary}

\vspace{1pc}

Due to this result we may {\it use the adjacent matrix of ${\vec G}(V^{0})$
as a working table} for determination MCP of the TCG ${\vec G}_{t}(V^{0})$
of a VS-digraph. Thus, we will suppose that chains of MPP of the TCG of
a VS-digraph are found using the adjacent matrix of this digraph. {\bf That
is, we choose only essential arcs to construct each new MPP!}

Certainly, we use the adjacent matrix of the transitive closure graph
to search for the maximum antichains $U(v)$ of such TCG.

The instance of constructing vertex-saturated digraph is shown 
in Appendix \ref{prm}.
\section{An algorithm for finding MMIS of a graph}
\label{MAXSET}

Let a saturated digraph ${\vec G}(V^{0})$ is constructed, which has 
a MMIS ${\hat U}$ such that $Card({\hat U})>Card(V^{0})$. In this case, at 
least one of the chains of TCG of the VS-digraph ${\vec G}(V^{0})$  
contains a fictitious arc, whose endpoints belong to the MMIS.

Let, further, a some fictitious arc $v_{i}\gg v_{j}$ is found in the TCG 
${\vec G}_{t}(V^{0})$. We shall remove the vertices $v_{i},v_{j}$ from
the digraph ${\vec G}_{t}(V^{0})$ and all vertices, which are adjacent with 
them. As a  result, we shall obtain a digraph 
${\vec G}_{1}(V_{1}^{0})=(V_{1},{\vec \Gamma}_{1})$, where
\[
V_{1}=V\setminus (\{v_{i},v_{j}\}\cup \Gamma(v_{i})\cup \Gamma(v_{j})),
\]
\[
V_{1}^{0}=V^{0}\setminus ({\vec \Gamma}^{-1} (v_{i})\cup {\vec \Gamma}^{-1}(v_{j})).\ \ 
{\vec \Gamma}_{1}={\vec \Gamma} \cap V_{1}.
\]
Here $\Gamma(v)={\vec \Gamma}(v)\cup {\vec \Gamma}^{-1}(v)$.

For the digraph ${\vec G}_{1}(V_{1}^{0})$, we shall use the procedure of 
constructing a VS-digraph by the algorithm VS. As a result, we shall
obtain a digraph ${\vec G}(Z^{0})$, which shall call {\it induced by 
removing the fictitious arc} $v_{i}\gg v_{j}$.

An algorithm for finding a MMIS of a digraph ${\vec G}(V^{0})$ is constructed 
on the supposition that the following conjecture is true.

\begin{conjecture}
\label{gip}
Let a saturated digraph ${\vec G}(V^{0})$ has an independent set $U\subset V$ 
such that $Card(U)>Card(V^{0})$. Then it will be found a fictitious arc 
$v_{i}\gg v_{j}$ such that in the digraph ${\vec G}(Z^{0})$, induced by 
removing this arc, the relation $Card(Z^{0})\geq Card(V^{0})-1$ is 
satisfied.
\end{conjecture}

The worded conjecture allows to formulate a solution algorithm for finding 
a MMIS of a graph $G\in L_{n}$. Input of the algorithm is an undirected 
graph $G\in L_{n}$. Output of the algorithm is the MMIS.

\begin{list}{}{
\setlength{\topsep}{2mm}
\setlength{\itemsep}{0mm}
\setlength{\parsep}{1mm}
}
\item
{\bf An algorithm for finding a MMIS.}
\item[{\it Step 1.}] Execute an initial orientation of the graph edges so 
to get an acyclic digraph $G(V^{0})$.
\item[{\it Step 2.}] Execute the algorithm VS for the digraph 
$G(V^{0})$.
\item[{\it Step 3.}] In TCG of the VS-digraph to find an unmarked 
fictitious arc $v_{i}\gg v_{j}$. Mark the found fictitious arc as considered.
If all fictitious arcs are marked, go to the Step 7.
\item[{\it Step 4.}] Remove vertices $v_{i},v_{j}$ as well as all adjacent 
with them vertices from the saturated digraph $G(V^{0})$. As a result,
a digraph ${\vec G}_{1}(V_{1}^{0})$ will be obtained.
\item[{\it Step 5.}] Execute the algorithm VS for the digraph 
$G_{1}(V_{1}^{0})$. As a result, a digraph ${\vec G}(Z^{0})$ will be obtained.
\item[{\it Step 6.}] If $Card(Z^{0})\geq Card(V^{0})-1$, construct a set 
$W=Z^{0}\cup \{v_{i},v_{j}\}$ and execute the cutting operation 
$\sigma_{W}({\vec G}(V^{0}))$ in the saturated digraph $G(V^{0})$. Go to
Step 2. Otherwise go back to Step 3.
\item[{\it Step 7.}] Put a MMIS ${\hat U}=V^{0}$.
\end{list}

\begin{theorem}
If the conjecture \ref{gip} is true then the stated algorithm 
finds a MMIS of the graph $G\in L_{n}$.
\end{theorem}

It is obviously.\hfill Q.E.D.

\begin{theorem}
The running time of the algorithm of finding a MMIS equals to $O(n^{8})$.
\end{theorem}

Indeed, single executing of the Steps 3 -- 6 requires $O(n_{1}^{5})$ of
time units, where $n_{1}$ is the number of vertices in the digraph 
${\vec G}(Z^{0})$, induced by removing a fictitious arc. Since total 
number of fictitious arcs is $O(n^{2})$, in worse case, for executing
the Steps 3 -- 6 is required $O(n^{7})$ of time units. If suppose that after 
executing these steps, the found independent set $V^{0}$ will be increased 
to the unit, the total running time of steps 2 -- 6 equals to 
$O(n^{8})$.\hfill Q.E.D.

\section{Conclusion}

The pascal-programs were written for the proposed algorithm. 
Long testing the program for random graphs has shown that the algorithm 
runs stably and correctly. 

Of course, the offered algorithm is not competitive in practice 
because of high degree of polynomial estimation of the running time.
However, the algorithm has important theoretical significance since 
there is a good probability to prove that for \np-complete problems 
possible to construct a polynomial-time algorithm.

Author thanks Guenter Stertenbrink\footnote{sktekkrtken(at)akol.com (remove k's)} for the help 
in testing the programs.


\newpage
\appendix{\LARGE\bf Appendix}
\section{Partially Ordered Sets}
\label{poset}

We recall some conceptions of Set Theory.

Relations between two objects are called {\it binary}. A binary relation $R$ 
may be represented by a listing of object pairs, which are in the relation $R$:
\begin{displaymath}
(a_{1}, b_{1}), \ldots , (a_{m}, b_{m}).
\end{displaymath}
If $(a, b)\in R$, then this fact we also denote by $aRb$. When 
$(a, b)\not\in R$,
then we will write $a{\bar {R}}b$. If $a\in A$ and $b\in A$ for all $aRb$,
then $R$ is called a {\it relation on the set $A$}. Further, we will
consider only relations on the {\it finite} set $A$.

If $A$ is a finite set and $R$ is a relation on $A$, we can represent $R$ 
as a digraph $\vec G$. Each element of $A$ is assigned to a vertex of
$\vec G$, and the vertex $a_{i}$ is joined with a vertex $a_{j}$ by the arc
$(a_{i}, a_{j})$ if and only if $a_{i} R a_{j}$.

A relation $R$ is {\it reflexive} if $aRa$ for every $a\in A$. A relation $R$
is {\bf irreflexive} if $a {\bar {R}} a$ for every $a\in A$.
A relation $R$ is {\it symmetric} if whenever $a R b$, then $b R a$. A
relation $R$ is {\it antisymmetric} if whenever $a R b$ and $b R a$, then
$a=b$. A relation $R$ is {\it transitive} if whenever $a R b$ and $b R c$,
then $a R c$.

A binary relation $R$ is called a {\it partial order} if $R$ is antisymmetric,
and transitive. The set $A$ together with the partial order $R$ are called a
{\it partially ordered set}. We will denote this partially ordered set by
$(A, R)$ or $(A, \geq)$. If a relation $R$ is irreflexive, then such partial
order is called {\it strict}. A strictly ordered set is written by $(A, >)$.

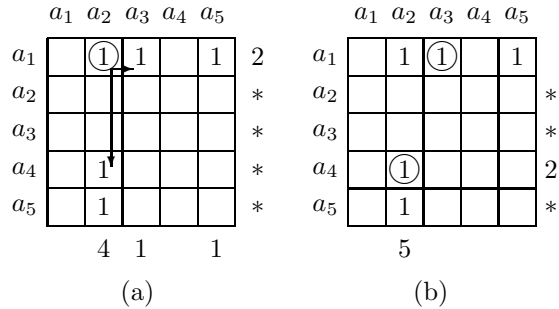
\begin{figure}[h]
\centering
\mbox{\unitlength=1.00mm
\special{em:linewidth 0.4pt}
\linethickness{0.4pt}
\begin{picture}(72.00,38.00)
\put(5.00,10.00){\line(0,1){25.00}}
\put(5.00,35.00){\line(1,0){25.00}}
\put(30.00,35.00){\line(0,-1){25.00}}
\put(30.00,10.00){\line(-1,0){25.00}}
\put(5.00,30.00){\line(1,0){25.00}}
\put(5.00,25.07){\line(1,0){25.00}}
\put(5.00,20.00){\line(1,0){25.00}}
\put(5.00,15.07){\line(1,0){25.00}}
\put(10.00,35.00){\line(0,-1){25.00}}
\put(15.00,35.00){\line(0,-1){25.00}}
\put(20.00,35.00){\line(0,-1){25.00}}
\put(25.00,35.00){\line(0,-1){25.00}}
\put(12.50,32.58){\makebox(0,0)[cc]{1}}
\put(12.50,32.60){\circle{3.90}}
\put(17.50,32.58){\makebox(0,0)[cc]{1}}
\put(27.50,32.58){\makebox(0,0)[cc]{1}}
\put(12.50,17.58){\makebox(0,0)[cc]{1}}
\put(12.50,12.58){\makebox(0,0)[cc]{1}}
\put(2.00,32.50){\makebox(0,0)[cc]{$a_{1}$}}
\put(2.00,27.50){\makebox(0,0)[cc]{$a_{2}$}}
\put(2.00,22.50){\makebox(0,0)[cc]{$a_{3}$}}
\put(2.00,17.50){\makebox(0,0)[cc]{$a_{4}$}}
\put(2.00,12.50){\makebox(0,0)[cc]{$a_{5}$}}
\put(7.00,38.00){\makebox(0,0)[cc]{$a_{1}$}}
\put(12.00,38.00){\makebox(0,0)[cc]{$a_{2}$}}
\put(17.07,38.00){\makebox(0,0)[cc]{$a_{3}$}}
\put(22.00,38.00){\makebox(0,0)[cc]{$a_{4}$}}
\put(27.07,38.00){\makebox(0,0)[cc]{$a_{5}$}}
\put(45.00,10.00){\line(0,1){25.00}}
\put(45.00,35.00){\line(1,0){25.00}}
\put(70.00,35.00){\line(0,-1){25.00}}
\put(70.00,10.00){\line(-1,0){25.00}}
\put(45.00,30.00){\line(1,0){25.00}}
\put(45.00,25.00){\line(1,0){25.00}}
\put(45.00,20.00){\line(1,0){25.00}}
\put(45.00,15.00){\line(1,0){25.00}}
\put(50.00,35.00){\line(0,-1){25.00}}
\put(55.00,35.00){\line(0,-1){25.00}}
\put(60.00,35.00){\line(0,-1){25.00}}
\put(65.00,35.00){\line(0,-1){25.00}}
\put(52.50,32.58){\makebox(0,0)[cc]{1}}
\put(57.50,32.58){\makebox(0,0)[cc]{1}}
\put(57.50,32.60){\circle{3.90}}
\put(67.50,32.58){\makebox(0,0)[cc]{1}}
\put(52.50,17.58){\makebox(0,0)[cc]{1}}
\put(52.50,17.60){\circle{3.90}}
\put(52.50,12.58){\makebox(0,0)[cc]{1}}
\put(42.00,32.50){\makebox(0,0)[cc]{$a_{1}$}}
\put(42.00,27.50){\makebox(0,0)[cc]{$a_{2}$}}
\put(42.00,22.50){\makebox(0,0)[cc]{$a_{3}$}}
\put(42.00,17.50){\makebox(0,0)[cc]{$a_{4}$}}
\put(42.00,12.50){\makebox(0,0)[cc]{$a_{5}$}}
\put(47.50,38.00){\makebox(0,0)[cc]{$a_{1}$}}
\put(52.50,38.00){\makebox(0,0)[cc]{$a_{2}$}}
\put(57.50,38.00){\makebox(0,0)[cc]{$a_{3}$}}
\put(62.50,38.00){\makebox(0,0)[cc]{$a_{4}$}}
\put(67.50,38.00){\makebox(0,0)[cc]{$a_{5}$}}
\put(33.00,27.50){\makebox(0,0)[cc]{$*$}}
\put(33.00,22.50){\makebox(0,0)[cc]{$*$}}
\put(33.00,17.50){\makebox(0,0)[cc]{$*$}}
\put(33.00,12.50){\makebox(0,0)[cc]{$*$}}
\put(17.50,7.00){\makebox(0,0)[cc]{$1$}}
\put(27.50,7.00){\makebox(0,0)[cc]{$1$}}
\put(12.50,7.00){\makebox(0,0)[cc]{$4$}}
\put(33.00,32.50){\makebox(0,0)[cc]{$2$}}
\put(72.00,27.50){\makebox(0,0)[cc]{$*$}}
\put(72.00,22.50){\makebox(0,0)[cc]{$*$}}
\put(72.00,17.50){\makebox(0,0)[cc]{$2$}}
\put(72.00,12.50){\makebox(0,0)[cc]{$*$}}
\put(52.50,7.00){\makebox(0,0)[cc]{$5$}}
\put(17.00,1.00){\makebox(0,0)[cc]{(a)}}
\put(56.00,1.00){\makebox(0,0)[cc]{(b)}}
\put(13.50,30.90){\vector(1,0){3.00}}
\put(13.50,30.90){\vector(0,-1){13.00}}
\end{picture}
}
\caption{The adjacent matrixes}
\label{f12}
\end{figure}

In Fig. \ref{f1-2}, the acyclic graph represents the partial order,
induced by the binary relation
\begin{displaymath}
R=\{(a_{1}, a_{2}), (a_{1}, a_{3}), (a_{1}, a_{5}), (a_{4}, a_{2}),
(a_{5}, a_{2})\}.
\end{displaymath}
Here and throughout, we assume that the orientation of arcs of a digraph on a
drawing is from below upwards.

Dilworth's famous theorem establishes the relationship between
a MCP and the maximum antichain of $(A, \geq)$ \cite{ford-fulkerson, west}.

\begin{theorem}
{\rm (Dilworth R.P.)} Let $(A, \geq)$ be a finite partially ordered set.
The minimum number of disjoint chains, which the set $(A, \geq)$ can be
partitioned on, equals to the capacity of the maximum antichain in $(A, \geq)$.
\end{theorem}

There is an efficient algorithm for the partitioning a finite partially
ordered set into the minimum number of chains and for finding the maximum 
antichain, elaborated by L. R. Ford and D. R. Fulkerson \cite{ford-fulkerson}.
In essence, this algorithm finds the maximum matching in a bipartite graph 
$G^{*}=(X,Y,\Gamma^{*})$. If a partially ordered set has $n$ elements, then 
this graph contains $2n$ vertices and $Card(X)=Card(Y)=n$. An edge 
$(x_{i}, y_{j})$ joins two vertices $x_{1}\in X$ and $y_{j}\in Y$ if and 
only if the corresponding elements $a_{1}, a_{2}\in A$ of $(A, \geq)$ are
comparable.

In manual computations, we will use an adjacent matrix $M$ of $G^{*}$
as a working table. Units of $M$ determine its {\it admissible cells}. 
Two cells of $M$ are called {\it independent} if they are located in distinct 
rows and distinct columns of $M$. To find the maximum matching, we will have
to find the maximum number of admissible independent cells of $M$.

The algorithm for partitioning a partially ordered set into the minimum
number of chains consists of two stages:
\begin{itemize}
\item Construct an initial partition of the partially ordered set into chains;
\item Improve the existing partition if it is possible.
\end{itemize}

To avoid ambiguity, we always look through rows and columns of $M$ uniformly:
from top to bottom in columns and from left to right in rows.

To obtain an initial partition, we may use the following procedure.

\begin{list}{}{
\setlength{\topsep}{2mm}
\setlength{\itemsep}{0mm}
\setlength{\parsep}{1mm}
}
\item[{\it Step 0.}] Put $N=n$, where $n$ is the number of rows $M$, $i=1$.
\item[{\it Step 1.}] If $N=0$, then complete the calculations as an initial 
partition is found.
\item[{\it Step 2.}] In $i$-th row of $M$ find the first on the order 
admissible cell, whose appropriate column is not marked. If such
cell is not found, put $i:=i+1$, $N:=N-1$ and go to Step 1. Otherwise, 
remember the found cell ($i, j$), mark a column $j$, calculate $i:=i+1$
and go to Step 1.
\end{list}

Fig. \ref{f12} (a) shows the adjacent matrix for the strictly ordered set,
represented in Fig. \ref{f1-2}. The chosen cells of the initial partition
are indicated by a circle.

To find a MCP and the maximum antichain of a set $(A, \geq)$, we will make
use of the Ford-Fulkerson's algorithm \cite{ford-fulkerson}. The algorithm
begins to work after termination of the previous procedure, that is, when
there exists an initial partition of the ordered set into chains.

\begin{list}{}{
\setlength{\topsep}{2mm}
\setlength{\itemsep}{0mm}
\setlength{\parsep}{1mm}
}
\item[{\it Step 1.}] Mark rows of $M$ that do not contain the chosen cells,
by the symbol $(*)$.
\item[{\it Step 2.}] Look through the newly marked rows of $M$ and find all
unchosen cells in each row. Mark all unmarked columns of $M$ that correspond
with such cells by an index of the row.
\item[{\it Step 3.}] Look through the newly marked columns. If an examined 
column contains a chosen cell (that is, the cell is enclosed within a circle), 
then mark the row containing the chosen cell by an index of the examined 
column. If the column does not contain a chosen cell, go to Step 4. If it is 
impossible to mark new rows, then go to Step 5.
\item[{\it Step 4.}] The essence of the given step is the procedure of
constructing a new collection of independent cells, each having one more
cell than the former collection. At each stage of this procedure, except for 
the final step, we pick a {\it new} admissible cell of $M$ and delete the 
``old'' one. Increasing the total number of chosen cells happens as follows.
In the found $j$-th column, choose a new cell in a row $m(j)$, where $m(j)$
is a mark of the current column. Let we already have chosen the cell ($i,j$), 
which marks $m(i)$ and $m(j)$ correspond to, where $m(i)$ is a mark  
the $i$-th row. If $m(j)=(*)$, the procedure of constructing a new collection 
of independent cells is completed. Delete all marks of rows and columns, and 
go to Step 2. Otherwise, delete the cell ($i,m(i)$) and choose a cell
($m(m(i))$, $m(i)$). Put $i=m(m(i))$, $j=m(i)$ and repeat the process 
described above.
\item[{\it Step 5.}] Find the maximum antichain $\cal U$ =
$U_{r}\setminus U_{c}$, where $U_{r}$ is a set of marked rows, and $U_{c}$
is a set of marked columns. Terminate the calculations. The found admissible 
cells determine arcs forming chains of the MCP.
\end{list}

Fig. \ref{f12} (b) shows the picked cells of the optimal partition for the
partially ordered set, represented in Fig. \ref{f1-2}. In this case, the
MCP consists of the chains $A_{1}=\{a_{1}, a_{3}\}$, $A_{2}=\{a_{4}, a_{2}\}$,
and $A_{3}=\{a_{5}\}$. We also have a set
$U_{r}=\{a_{2}, a_{3}, a_{4}, a_{5}\}$
of marked rows, and a set $U_{c}=\{a_{2}\}$ of marked columns. Consequently,
the maximum antichain $\cal U$ is equal to
\begin{displaymath}
{\cal U}= \{a_{2}, a_{3}, a_{4}, a_{5}\}\setminus \{a_{2}\} = \{a_{3},
a_{4}, a_{5}\}.
\end{displaymath}

\vspace{1pc}

Ford-Fulkerson's methodology, described above, for finding antichains may
be easily adapted to any algorithm of finding the maximum matching in a
bipartite graph, for example, Hopcroft-Karp's algorithm
\cite{hopcroft-karp}, \cite{swamy-thulasiraman}, or flow algorithm
\cite{papadimitriou-steiglitz}. Therefore, we assume that the running-time
of a MCP construction is equal to $O(n^{5/2})$.

\section{An instance of constructing a vertex-saturated digraph}
\label{prm}

Consider an instance of constructing a vertex-saturated digraph (VS-digraph).

\vspace{1pc}

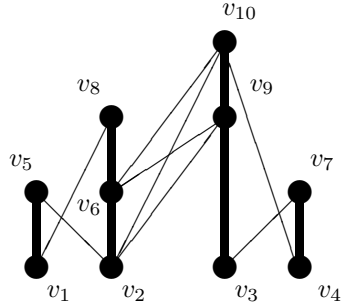
\begin{figure}[hbp]
\centering
\mbox{\unitlength=1mm
\special{em:linewidth 0.4pt}
\linethickness{0.4pt}
\begin{picture}(44.00,39.00)
\put(5.00,15.00){\line(1,-1){10.00}}
\put(15.00,25.00){\line(-1,-2){10.00}}
\put(15.00,15.00){\line(3,4){15.00}}
\put(30.13,35.00){\line(-1,-2){15.00}}
\put(15.00,5.00){\line(3,4){15.00}}
\put(30.13,25.00){\line(-3,-2){15.00}}
\put(30.00,5.00){\line(1,1){10.00}}
\put(40.00,4.00){\line(-1,3){10.40}}
\put(5.00,5.00){\circle*{3.00}}
\put(15.00,5.00){\circle*{3.00}}
\put(30.00,5.00){\circle*{3.00}}
\put(40.00,5.00){\circle*{3.00}}
\put(5.00,15.00){\circle*{3.00}}
\put(15.00,15.00){\circle*{3.00}}
\put(40.00,15.00){\circle*{3.00}}
\put(15.00,25.00){\circle*{3.00}}
\put(30.00,25.00){\circle*{3.00}}
\put(30.00,35.00){\circle*{3.00}}
\put(8.00,1.50){\makebox(0,0)[cc]{$v_{1}$}}
\put(18.00,1.50){\makebox(0,0)[cc]{$v_{2}$}}
\put(33.00,1.50){\makebox(0,0)[cc]{$v_{3}$}}
\put(44.00,1.50){\makebox(0,0)[cc]{$v_{4}$}}
\put(3.00,19.00){\makebox(0,0)[cc]{$v_{5}$}}
\put(12.00,13.00){\makebox(0,0)[cc]{$v_{6}$}}
\put(43.00,19.00){\makebox(0,0)[cc]{$v_{7}$}}
\put(12.00,29.00){\makebox(0,0)[cc]{$v_{8}$}}
\put(35.00,29.00){\makebox(0,0)[cc]{$v_{9}$}}
\put(32.00,39.00){\makebox(0,0)[cc]{$v_{10}$}}
\linethickness{3.0pt}
\put(5.00,5.00){\line(0,1){10.00}}
\put(15.00,5.00){\line(0,1){20.00}}
\put(30.00,35.00){\line(0,-1){30.00}}
\put(40.00,15.00){\line(0,-1){11.00}}
\end{picture}
}
\caption{A digraph}
\label{f2-4}
\end{figure}

Fig. \ref{f2-4} shows a digraph ${\vec G}(X^{0})$, obtained from an initial 
undirected graph $G$ as it was described in part \ref{init}. Recall that 
the orientation of arcs of the digraph in figures is from below upwards.

Construct a VS-digraph.

\begin{figure}[hbp]
\centering
\mbox{\unitlength 1.00mm
\special{em:linewidth 0.4pt}
\linethickness{0.4pt}
\begin{picture}(122.50,189.75)
\put(5.00,73.00){\line(0,1){50.00}}
\put(5.00,123.00){\line(1,0){50.00}}
\put(55.00,123.00){\line(0,-1){50.00}}
\put(55.00,73.00){\line(-1,0){50.00}}
\put(5.00,118.00){\line(1,0){50.00}}
\put(5.00,113.00){\line(1,0){50.00}}
\put(5.00,108.00){\line(1,0){50.00}}
\put(5.00,103.00){\line(1,0){50.00}}
\put(5.00,98.00){\line(1,0){50.00}}
\put(5.00,93.00){\line(1,0){50.00}}
\put(5.00,88.00){\line(1,0){50.00}}
\put(5.00,83.00){\line(1,0){50.00}}
\put(5.00,78.00){\line(1,0){50.00}}
\put(10.00,123.00){\line(0,-1){50.00}}
\put(15.00,123.00){\line(0,-1){50.00}}
\put(20.00,123.00){\line(0,-1){50.00}}
\put(25.00,123.00){\line(0,-1){50.00}}
\put(30.00,123.00){\line(0,-1){50.00}}
\put(35.00,123.00){\line(0,-1){50.00}}
\put(40.00,123.00){\line(0,-1){50.00}}
\put(45.00,123.00){\line(0,-1){50.00}}
\put(50.00,123.00){\line(0,-1){50.00}}
\put(1.00,120.50){\makebox(0,0)[cc]{$v_{1}$}}
\put(1.00,115.50){\makebox(0,0)[cc]{$v_{2}$}}
\put(1.00,110.50){\makebox(0,0)[cc]{$v_{3}$}}
\put(1.00,105.50){\makebox(0,0)[cc]{$v_{4}$}}
\put(1.00,100.50){\makebox(0,0)[cc]{$v_{5}$}}
\put(1.00,95.50){\makebox(0,0)[cc]{$v_{6}$}}
\put(1.00,90.50){\makebox(0,0)[cc]{$v_{7}$}}
\put(1.00,85.50){\makebox(0,0)[cc]{$v_{8}$}}
\put(1.00,80.50){\makebox(0,0)[cc]{$v_{9}$}}
\put(1.00,75.50){\makebox(0,0)[cc]{$v_{10}$}}
\put(7.00,125.75){\makebox(0,0)[cc]{$v_{1}$}}
\put(12.00,125.75){\makebox(0,0)[cc]{$v_{2}$}}
\put(17.00,125.75){\makebox(0,0)[cc]{$v_{3}$}}
\put(22.00,125.75){\makebox(0,0)[cc]{$v_{4}$}}
\put(27.00,125.75){\makebox(0,0)[cc]{$v_{5}$}}
\put(32.00,125.75){\makebox(0,0)[cc]{$v_{6}$}}
\put(37.00,125.75){\makebox(0,0)[cc]{$v_{7}$}}
\put(42.00,125.75){\makebox(0,0)[cc]{$v_{8}$}}
\put(47.00,125.75){\makebox(0,0)[cc]{$v_{9}$}}
\put(52.00,125.75){\makebox(0,0)[cc]{$v_{10}$}}
\put(27.50,120.50){\makebox(0,0)[cc]{1}}
\put(27.50,120.52){\circle{3.90}}
\put(42.50,120.50){\makebox(0,0)[cc]{1}}
\put(27.50,115.50){\makebox(0,0)[cc]{1}}
\put(32.50,115.50){\makebox(0,0)[cc]{1}}
\put(32.50,115.52){\circle{3.90}}
\put(42.50,115.50){\makebox(0,0)[cc]{f}}
\put(47.50,115.50){\makebox(0,0)[cc]{1}}
\put(52.50,115.50){\makebox(0,0)[cc]{1}}
\put(37.50,110.50){\makebox(0,0)[cc]{1}}
\put(47.50,110.50){\makebox(0,0)[cc]{1}}
\put(47.50,110.52){\circle{3.90}}
\put(52.50,110.50){\makebox(0,0)[cc]{f}}
\put(37.50,105.50){\makebox(0,0)[cc]{1}}
\put(37.50,105.52){\circle{3.90}}
\put(52.50,105.50){\makebox(0,0)[cc]{1}}
\put(42.50,95.50){\makebox(0,0)[cc]{1}}
\put(42.50,95.52){\circle{3.90}}
\put(47.50,95.50){\makebox(0,0)[cc]{1}}
\put(52.50,95.50){\makebox(0,0)[cc]{1}}
\put(52.50,80.50){\makebox(0,0)[cc]{1}}
\put(52.50,80.52){\circle{3.90}}
\put(57.50,120.50){\makebox(0,0)[cc]{5}}
\put(57.50,115.50){\makebox(0,0)[cc]{$*$}}
\put(57.50,110.50){\makebox(0,0)[cc]{9}}
\put(57.50,105.50){\makebox(0,0)[cc]{7}}
\put(57.50,100.50){\makebox(0,0)[cc]{$*$}}
\put(57.50,95.50){\makebox(0,0)[cc]{8}}
\put(57.50,90.50){\makebox(0,0)[cc]{$*$}}
\put(57.50,85.50){\makebox(0,0)[cc]{$*$}}
\put(57.50,80.50){\makebox(0,0)[cc]{10}}%
\put(57.50,75.50){\makebox(0,0)[cc]{$*$}}
\put(52.50,70.50){\makebox(0,0)[cc]{2}}
\put(47.50,70.50){\makebox(0,0)[cc]{2}}
\put(42.50,70.50){\makebox(0,0)[cc]{2}}
\put(37.50,70.50){\makebox(0,0)[cc]{3}}
\put(32.50,70.50){\makebox(0,0)[cc]{2}}
\put(27.50,70.50){\makebox(0,0)[cc]{2}}
\put(27.50,65.50){\makebox(0,0)[cc]{(c)}}
\put(70.00,73.00){\line(0,1){50.00}}
\put(70.00,123.00){\line(1,0){50.00}}
\put(120.00,123.00){\line(0,-1){50.00}}
\put(120.00,73.00){\line(-1,0){50.00}}
\put(70.00,118.00){\line(1,0){50.00}}
\put(70.00,113.00){\line(1,0){50.00}}
\put(70.00,108.00){\line(1,0){50.00}}
\put(70.00,103.00){\line(1,0){50.00}}
\put(70.00,98.00){\line(1,0){50.00}}
\put(70.00,93.00){\line(1,0){50.00}}
\put(70.00,88.00){\line(1,0){50.00}}
\put(70.00,83.00){\line(1,0){50.00}}
\put(70.00,78.00){\line(1,0){50.00}}
\put(75.00,123.00){\line(0,-1){50.00}}
\put(80.00,123.00){\line(0,-1){50.00}}
\put(85.00,123.00){\line(0,-1){50.00}}
\put(90.00,123.00){\line(0,-1){50.00}}
\put(95.00,123.00){\line(0,-1){50.00}}
\put(100.00,123.00){\line(0,-1){50.00}}
\put(105.00,123.00){\line(0,-1){50.00}}
\put(110.00,123.00){\line(0,-1){50.00}}
\put(115.00,123.00){\line(0,-1){50.00}}
\put(66.00,120.50){\makebox(0,0)[cc]{$v_{1}$}}
\put(66.00,115.50){\makebox(0,0)[cc]{$v_{2}$}}
\put(66.00,110.50){\makebox(0,0)[cc]{$v_{3}$}}
\put(66.00,105.50){\makebox(0,0)[cc]{$v_{4}$}}
\put(66.00,100.50){\makebox(0,0)[cc]{$v_{5}$}}
\put(66.00,95.50){\makebox(0,0)[cc]{$v_{6}$}}
\put(66.00,90.50){\makebox(0,0)[cc]{$v_{7}$}}
\put(66.00,85.50){\makebox(0,0)[cc]{$v_{8}$}}
\put(66.00,80.50){\makebox(0,0)[cc]{$v_{9}$}}
\put(66.00,75.50){\makebox(0,0)[cc]{$v_{10}$}}
\put(72.00,125.75){\makebox(0,0)[cc]{$v_{1}$}}
\put(77.00,125.75){\makebox(0,0)[cc]{$v_{2}$}}
\put(82.00,125.75){\makebox(0,0)[cc]{$v_{3}$}}
\put(87.00,125.75){\makebox(0,0)[cc]{$v_{4}$}}
\put(92.00,125.75){\makebox(0,0)[cc]{$v_{5}$}}
\put(97.00,125.75){\makebox(0,0)[cc]{$v_{6}$}}
\put(102.00,125.75){\makebox(0,0)[cc]{$v_{7}$}}
\put(107.00,125.75){\makebox(0,0)[cc]{$v_{8}$}}
\put(112.00,125.75){\makebox(0,0)[cc]{$v_{9}$}}
\put(117.00,125.75){\makebox(0,0)[cc]{$v_{10}$}}
\put(92.50,120.50){\makebox(0,0)[cc]{1}}
\put(92.50,120.52){\circle{3.90}}
\put(107.50,120.50){\makebox(0,0)[cc]{1}}
\put(92.50,115.50){\makebox(0,0)[cc]{1}}
\put(97.50,115.50){\makebox(0,0)[cc]{1}}
\put(97.50,115.52){\circle{3.90}}
\put(107.50,115.50){\makebox(0,0)[cc]{f}}
\put(107.50,115.50){\makebox(0,0)[cc]{1}}
\put(117.50,115.50){\makebox(0,0)[cc]{1}}
\put(102.50,110.50){\makebox(0,0)[cc]{1}}
\put(112.50,110.50){\makebox(0,0)[cc]{1}}
\put(112.50,110.52){\circle{3.90}}
\put(117.50,110.50){\makebox(0,0)[cc]{f}}
\put(102.50,105.50){\makebox(0,0)[cc]{1}}
\put(102.50,105.52){\circle{3.90}}
\put(117.50,105.50){\makebox(0,0)[cc]{1}}
\put(107.50,95.50){\makebox(0,0)[cc]{1}}
\put(107.50,95.52){\circle{3.90}}
\put(112.50,95.50){\makebox(0,0)[cc]{1}}
\put(117.50,95.50){\makebox(0,0)[cc]{1}}
\put(117.50,80.50){\makebox(0,0)[cc]{1}}
\put(117.50,80.52){\circle{3.90}}
\put(122.50,110.50){\makebox(0,0)[cc]{$*$}}
\put(122.50,105.50){\makebox(0,0)[cc]{7}}
\put(122.50,100.50){\makebox(0,0)[cc]{$*$}}
\put(122.50,90.50){\makebox(0,0)[cc]{$*$}}
\put(122.50,85.50){\makebox(0,0)[cc]{$*$}}
\put(122.50,80.50){\makebox(0,0)[cc]{10}}
\put(122.50,75.50){\makebox(0,0)[cc]{$*$}}
\put(117.50,70.50){\makebox(0,0)[cc]{3}}
\put(112.50,70.50){\makebox(0,0)[cc]{3}}
\put(102.50,70.50){\makebox(0,0)[cc]{3}}
\put(92.50,65.50){\makebox(0,0)[cc]{(d)}}%
\put(5.00,137.00){\line(0,1){50.00}}
\put(5.00,187.00){\line(1,0){50.00}}
\put(55.00,187.00){\line(0,-1){50.00}}
\put(55.00,137.00){\line(-1,0){50.00}}
\put(5.00,182.00){\line(1,0){50.00}}
\put(5.00,177.00){\line(1,0){50.00}}
\put(5.00,172.00){\line(1,0){50.00}}
\put(5.00,167.00){\line(1,0){50.00}}
\put(5.00,162.00){\line(1,0){50.00}}
\put(5.00,157.00){\line(1,0){50.00}}
\put(5.00,152.00){\line(1,0){50.00}}
\put(5.00,147.00){\line(1,0){50.00}}
\put(5.00,142.00){\line(1,0){50.00}}
\put(10.00,187.00){\line(0,-1){50.00}}
\put(15.00,187.00){\line(0,-1){50.00}}
\put(20.00,187.00){\line(0,-1){50.00}}
\put(25.00,187.00){\line(0,-1){50.00}}
\put(30.00,187.00){\line(0,-1){50.00}}
\put(35.00,187.00){\line(0,-1){50.00}}
\put(40.00,187.00){\line(0,-1){50.00}}
\put(45.00,187.00){\line(0,-1){50.00}}
\put(50.00,187.00){\line(0,-1){50.00}}
\put(1.00,184.50){\makebox(0,0)[cc]{$v_{1}$}}
\put(1.00,179.50){\makebox(0,0)[cc]{$v_{2}$}}
\put(1.00,174.50){\makebox(0,0)[cc]{$v_{3}$}}
\put(1.00,169.50){\makebox(0,0)[cc]{$v_{4}$}}
\put(1.00,174.50){\makebox(0,0)[cc]{$v_{5}$}}
\put(1.00,159.50){\makebox(0,0)[cc]{$v_{6}$}}
\put(1.00,154.50){\makebox(0,0)[cc]{$v_{7}$}}
\put(1.00,149.50){\makebox(0,0)[cc]{$v_{8}$}}
\put(1.00,144.50){\makebox(0,0)[cc]{$v_{9}$}}
\put(1.00,139.50){\makebox(0,0)[cc]{$v_{10}$}}
\put(7.00,189.75){\makebox(0,0)[cc]{$v_{1}$}}
\put(12.00,189.75){\makebox(0,0)[cc]{$v_{2}$}}
\put(17.00,189.75){\makebox(0,0)[cc]{$v_{3}$}}
\put(22.00,189.75){\makebox(0,0)[cc]{$v_{4}$}}
\put(27.00,189.75){\makebox(0,0)[cc]{$v_{5}$}}
\put(32.00,189.75){\makebox(0,0)[cc]{$v_{6}$}}
\put(37.00,189.75){\makebox(0,0)[cc]{$v_{7}$}}
\put(42.00,189.75){\makebox(0,0)[cc]{$v_{8}$}}
\put(47.00,189.75){\makebox(0,0)[cc]{$v_{9}$}}
\put(52.00,189.75){\makebox(0,0)[cc]{$v_{10}$}}
\put(27.50,184.50){\makebox(0,0)[cc]{1}}
\put(27.50,184.52){\circle{3.90}}
\put(42.50,184.50){\makebox(0,0)[cc]{1}}
\put(27.50,179.50){\makebox(0,0)[cc]{1}}
\put(32.50,179.50){\makebox(0,0)[cc]{1}}
\put(32.50,179.52){\circle{3.90}}
\put(42.50,179.50){\makebox(0,0)[cc]{f}}
\put(47.50,179.50){\makebox(0,0)[cc]{1}}
\put(52.50,179.50){\makebox(0,0)[cc]{1}}
\put(37.50,174.50){\makebox(0,0)[cc]{1}}
\put(47.50,174.50){\makebox(0,0)[cc]{1}}
\put(47.50,174.52){\circle{3.90}}
\put(52.50,174.50){\makebox(0,0)[cc]{f}}
\put(37.50,169.50){\makebox(0,0)[cc]{1}}
\put(37.50,169.52){\circle{3.90}}
\put(52.50,169.50){\makebox(0,0)[cc]{1}}
\put(42.50,159.50){\makebox(0,0)[cc]{1}}
\put(42.50,159.52){\circle{3.90}}
\put(47.50,159.50){\makebox(0,0)[cc]{1}}
\put(52.50,159.50){\makebox(0,0)[cc]{1}}
\put(52.50,144.50){\makebox(0,0)[cc]{1}}
\put(52.50,144.52){\circle{3.90}}
\put(57.50,164.50){\makebox(0,0)[cc]{$*$}}
\put(57.50,154.50){\makebox(0,0)[cc]{$*$}}
\put(57.50,149.50){\makebox(0,0)[cc]{$*$}}
\put(57.50,139.50){\makebox(0,0)[cc]{$*$}}
\put(27.50,129.50){\makebox(0,0)[cc]{(a)}}

\put(70.00,137.00){\line(0,1){50.00}}
\put(70.00,187.00){\line(1,0){50.00}}
\put(120.00,187.00){\line(0,-1){50.00}}
\put(120.00,137.00){\line(-1,0){50.00}}
\put(70.00,182.00){\line(1,0){50.00}}
\put(70.00,177.00){\line(1,0){50.00}}
\put(70.00,172.00){\line(1,0){50.00}}
\put(70.00,167.00){\line(1,0){50.00}}
\put(70.00,162.00){\line(1,0){50.00}}
\put(70.00,157.00){\line(1,0){50.00}}
\put(70.00,152.00){\line(1,0){50.00}}
\put(70.00,147.00){\line(1,0){50.00}}
\put(70.00,142.00){\line(1,0){50.00}}
\put(75.00,187.00){\line(0,-1){50.00}}
\put(80.00,187.00){\line(0,-1){50.00}}
\put(85.00,187.00){\line(0,-1){50.00}}
\put(90.00,187.00){\line(0,-1){50.00}}
\put(95.00,187.00){\line(0,-1){50.00}}
\put(100.00,187.00){\line(0,-1){50.00}}
\put(105.00,187.00){\line(0,-1){50.00}}
\put(110.00,187.00){\line(0,-1){50.00}}
\put(115.00,187.00){\line(0,-1){50.00}}
\put(66.00,184.50){\makebox(0,0)[cc]{$v_{1}$}}
\put(66.00,179.50){\makebox(0,0)[cc]{$v_{2}$}}
\put(66.00,174.50){\makebox(0,0)[cc]{$v_{3}$}}
\put(66.00,169.50){\makebox(0,0)[cc]{$v_{4}$}}
\put(66.00,164.50){\makebox(0,0)[cc]{$v_{5}$}}
\put(66.00,159.50){\makebox(0,0)[cc]{$v_{6}$}}
\put(66.00,154.50){\makebox(0,0)[cc]{$v_{7}$}}
\put(66.00,149.50){\makebox(0,0)[cc]{$v_{8}$}}
\put(66.00,144.50){\makebox(0,0)[cc]{$v_{9}$}}
\put(66.00,139.50){\makebox(0,0)[cc]{$v_{10}$}}
\put(72.00,189.75){\makebox(0,0)[cc]{$v_{1}$}}
\put(77.00,189.75){\makebox(0,0)[cc]{$v_{2}$}}
\put(82.00,189.75){\makebox(0,0)[cc]{$v_{3}$}}
\put(87.00,189.75){\makebox(0,0)[cc]{$v_{4}$}}
\put(92.00,189.75){\makebox(0,0)[cc]{$v_{5}$}}
\put(97.00,189.75){\makebox(0,0)[cc]{$v_{6}$}}
\put(102.00,189.75){\makebox(0,0)[cc]{$v_{7}$}}
\put(107.00,189.75){\makebox(0,0)[cc]{$v_{8}$}}
\put(112.00,189.75){\makebox(0,0)[cc]{$v_{9}$}}
\put(117.00,189.75){\makebox(0,0)[cc]{$v_{10}$}}
\put(92.50,184.50){\makebox(0,0)[cc]{1}}
\put(92.50,184.52){\circle{3.90}}
\put(107.50,184.50){\makebox(0,0)[cc]{1}}
\put(92.50,179.50){\makebox(0,0)[cc]{1}}
\put(97.50,179.50){\makebox(0,0)[cc]{1}}
\put(97.50,179.52){\circle{3.90}}
\put(107.50,179.50){\makebox(0,0)[cc]{f}}
\put(112.50,179.50){\makebox(0,0)[cc]{1}}
\put(117.50,179.50){\makebox(0,0)[cc]{1}}
\put(102.50,174.50){\makebox(0,0)[cc]{1}}
\put(112.50,174.50){\makebox(0,0)[cc]{1}}
\put(112.50,174.52){\circle{3.90}}
\put(117.50,174.50){\makebox(0,0)[cc]{f}}
\put(102.50,169.50){\makebox(0,0)[cc]{1}}
\put(102.50,169.52){\circle{3.90}}
\put(117.50,169.50){\makebox(0,0)[cc]{1}}
\put(107.50,159.50){\makebox(0,0)[cc]{1}}
\put(107.50,159.52){\circle{3.90}}
\put(112.50,159.50){\makebox(0,0)[cc]{1}}
\put(117.50,159.50){\makebox(0,0)[cc]{1}}
\put(117.50,144.50){\makebox(0,0)[cc]{1}}
\put(117.50,144.52){\circle{3.90}}
\put(122.50,184.50){\makebox(0,0)[cc]{$*$}}
\put(122.50,174.50){\makebox(0,0)[cc]{9}}
\put(122.50,169.50){\makebox(0,0)[cc]{7}}
\put(122.50,164.50){\makebox(0,0)[cc]{$*$}}
\put(122.50,159.50){\makebox(0,0)[cc]{8}}
\put(122.50,154.50){\makebox(0,0)[cc]{$*$}}
\put(122.50,149.50){\makebox(0,0)[cc]{$*$}}
\put(122.50,144.50){\makebox(0,0)[cc]{10}}
\put(122.50,139.50){\makebox(0,0)[cc]{$*$}}
\put(117.50,134.50){\makebox(0,0)[cc]{6}}
\put(112.50,134.50){\makebox(0,0)[cc]{6}}
\put(107.50,134.50){\makebox(0,0)[cc]{1}}
\put(102.50,134.50){\makebox(0,0)[cc]{3}}
\put(92.50,134.50){\makebox(0,0)[cc]{1}}
\put(92.50,129.50){\makebox(0,0)[cc]{(b)}}

\put(5.00,9.00){\line(0,1){50.00}}
\put(5.00,59.00){\line(1,0){50.00}}
\put(55.00,59.00){\line(0,-1){50.00}}
\put(55.00,9.00){\line(-1,0){50.00}}
\put(5.00,54.00){\line(1,0){50.00}}
\put(5.00,49.00){\line(1,0){50.00}}
\put(5.00,44.00){\line(1,0){50.00}}
\put(5.00,39.00){\line(1,0){50.00}}
\put(5.00,34.00){\line(1,0){50.00}}
\put(5.00,29.00){\line(1,0){50.00}}
\put(5.00,24.00){\line(1,0){50.00}}
\put(5.00,19.00){\line(1,0){50.00}}
\put(5.00,14.00){\line(1,0){50.00}}
\put(10.00,59.00){\line(0,-1){50.00}}
\put(15.00,59.00){\line(0,-1){50.00}}
\put(20.00,59.00){\line(0,-1){50.00}}
\put(25.00,59.00){\line(0,-1){50.00}}
\put(30.00,59.00){\line(0,-1){50.00}}
\put(35.00,59.00){\line(0,-1){50.00}}
\put(40.00,59.00){\line(0,-1){50.00}}
\put(45.00,59.00){\line(0,-1){50.00}}
\put(50.00,59.00){\line(0,-1){50.00}}
\put(1.00,56.50){\makebox(0,0)[cc]{$v_{1}$}}
\put(1.00,51.50){\makebox(0,0)[cc]{$v_{2}$}}
\put(1.00,46.50){\makebox(0,0)[cc]{$v_{3}$}}
\put(1.00,41.50){\makebox(0,0)[cc]{$v_{4}$}}
\put(1.00,36.50){\makebox(0,0)[cc]{$v_{5}$}}
\put(1.00,31.50){\makebox(0,0)[cc]{$v_{6}$}}
\put(1.00,26.50){\makebox(0,0)[cc]{$v_{7}$}}
\put(1.00,21.50){\makebox(0,0)[cc]{$v_{8}$}}
\put(1.00,16.50){\makebox(0,0)[cc]{$v_{9}$}}
\put(1.00,11.50){\makebox(0,0)[cc]{$v_{10}$}}
\put(7.00,61.75){\makebox(0,0)[cc]{$v_{1}$}}
\put(12.00,61.75){\makebox(0,0)[cc]{$v_{2}$}}
\put(17.00,61.75){\makebox(0,0)[cc]{$v_{3}$}}
\put(22.00,61.75){\makebox(0,0)[cc]{$v_{4}$}}
\put(27.00,61.75){\makebox(0,0)[cc]{$v_{5}$}}
\put(32.00,61.75){\makebox(0,0)[cc]{$v_{6}$}}
\put(37.00,61.75){\makebox(0,0)[cc]{$v_{7}$}}
\put(42.00,61.75){\makebox(0,0)[cc]{$v_{8}$}}
\put(47.00,61.75){\makebox(0,0)[cc]{$v_{9}$}}
\put(52.00,61.75){\makebox(0,0)[cc]{$v_{10}$}}
\put(27.50,56.50){\makebox(0,0)[cc]{1}}
\put(27.50,56.52){\circle{3.90}}
\put(42.50,56.50){\makebox(0,0)[cc]{1}}
\put(27.50,51.50){\makebox(0,0)[cc]{1}}
\put(32.50,51.50){\makebox(0,0)[cc]{1}}
\put(32.50,51.52){\circle{3.90}}
\put(42.50,51.50){\makebox(0,0)[cc]{f}}
\put(47.50,51.50){\makebox(0,0)[cc]{1}}
\put(52.50,51.50){\makebox(0,0)[cc]{1}}
\put(37.50,46.50){\makebox(0,0)[cc]{1}}
\put(47.50,46.50){\makebox(0,0)[cc]{1}}
\put(47.50,46.52){\circle{3.90}}
\put(52.50,46.50){\makebox(0,0)[cc]{f}}
\put(37.50,41.50){\makebox(0,0)[cc]{1}}
\put(37.50,41.52){\circle{3.90}}
\put(52.50,41.50){\makebox(0,0)[cc]{1}}
\put(42.50,31.50){\makebox(0,0)[cc]{1}}
\put(42.50,31.52){\circle{3.90}}
\put(47.50,31.50){\makebox(0,0)[cc]{1}}
\put(52.50,31.50){\makebox(0,0)[cc]{1}}
\put(52.50,16.50){\makebox(0,0)[cc]{1}}
\put(52.50,16.52){\circle{3.90}}
\put(57.50,41.50){\makebox(0,0)[cc]{$*$}}
\put(57.50,36.50){\makebox(0,0)[cc]{$*$}}
\put(57.50,26.50){\makebox(0,0)[cc]{$*$}}
\put(57.50,21.50){\makebox(0,0)[cc]{$*$}}
\put(57.50,16.50){\makebox(0,0)[cc]{10}}
\put(57.50,11.50){\makebox(0,0)[cc]{$*$}}
\put(52.50,6.50){\makebox(0,0)[cc]{4}}
\put(37.50,6.50){\makebox(0,0)[cc]{4}}
\put(27.50,1.50){\makebox(0,0)[cc]{(e)}}

\put(70.00,9.00){\line(0,1){50.00}}
\put(70.00,59.00){\line(1,0){50.00}}
\put(120.00,59.00){\line(0,-1){50.00}}
\put(120.00,9.00){\line(-1,0){50.00}}
\put(70.00,54.00){\line(1,0){50.00}}
\put(70.00,49.00){\line(1,0){50.00}}
\put(70.00,44.00){\line(1,0){50.00}}
\put(70.00,39.00){\line(1,0){50.00}}
\put(70.00,34.00){\line(1,0){50.00}}
\put(70.00,29.00){\line(1,0){50.00}}
\put(70.00,24.00){\line(1,0){50.00}}
\put(70.00,19.00){\line(1,0){50.00}}
\put(70.00,14.00){\line(1,0){50.00}}
\put(75.00,59.00){\line(0,-1){50.00}}
\put(80.00,59.00){\line(0,-1){50.00}}
\put(85.00,59.00){\line(0,-1){50.00}}
\put(90.00,59.00){\line(0,-1){50.00}}
\put(95.00,59.00){\line(0,-1){50.00}}
\put(100.00,59.00){\line(0,-1){50.00}}
\put(105.00,59.00){\line(0,-1){50.00}}
\put(110.00,59.00){\line(0,-1){50.00}}
\put(115.00,59.00){\line(0,-1){50.00}}
\put(66.00,56.50){\makebox(0,0)[cc]{$v_{1}$}}
\put(66.00,51.50){\makebox(0,0)[cc]{$v_{2}$}}
\put(66.00,46.50){\makebox(0,0)[cc]{$v_{3}$}}
\put(66.00,41.50){\makebox(0,0)[cc]{$v_{4}$}}
\put(66.00,36.50){\makebox(0,0)[cc]{$v_{5}$}}
\put(66.00,31.50){\makebox(0,0)[cc]{$v_{6}$}}
\put(66.00,26.50){\makebox(0,0)[cc]{$v_{7}$}}
\put(66.00,21.50){\makebox(0,0)[cc]{$v_{8}$}}
\put(66.00,16.50){\makebox(0,0)[cc]{$v_{9}$}}
\put(66.00,11.50){\makebox(0,0)[cc]{$v_{10}$}}
\put(72.00,61.75){\makebox(0,0)[cc]{$v_{1}$}}
\put(77.00,61.75){\makebox(0,0)[cc]{$v_{2}$}}
\put(82.00,61.75){\makebox(0,0)[cc]{$v_{3}$}}
\put(87.00,61.75){\makebox(0,0)[cc]{$v_{4}$}}
\put(92.00,61.75){\makebox(0,0)[cc]{$v_{5}$}}
\put(97.00,61.75){\makebox(0,0)[cc]{$v_{6}$}}
\put(102.00,61.75){\makebox(0,0)[cc]{$v_{7}$}}
\put(107.00,61.75){\makebox(0,0)[cc]{$v_{8}$}}
\put(112.00,61.75){\makebox(0,0)[cc]{$v_{9}$}}
\put(117.00,61.75){\makebox(0,0)[cc]{$v_{10}$}}
\put(92.50,56.50){\makebox(0,0)[cc]{1}}
\put(92.50,56.52){\circle{3.90}}
\put(107.50,56.50){\makebox(0,0)[cc]{1}}
\put(92.50,51.50){\makebox(0,0)[cc]{1}}
\put(97.50,51.50){\makebox(0,0)[cc]{1}}
\put(97.50,51.52){\circle{3.90}}
\put(107.50,51.50){\makebox(0,0)[cc]{f}}
\put(112.50,51.50){\makebox(0,0)[cc]{1}}
\put(117.50,51.50){\makebox(0,0)[cc]{1}}
\put(102.50,46.50){\makebox(0,0)[cc]{1}}
\put(112.50,46.50){\makebox(0,0)[cc]{1}}
\put(112.50,46.52){\circle{3.90}}
\put(117.50,46.50){\makebox(0,0)[cc]{f}}
\put(102.50,41.50){\makebox(0,0)[cc]{1}}
\put(102.50,41.52){\circle{3.90}}
\put(117.50,41.50){\makebox(0,0)[cc]{1}}
\put(107.50,31.50){\makebox(0,0)[cc]{1}}
\put(107.50,31.52){\circle{3.90}}
\put(112.50,31.50){\makebox(0,0)[cc]{1}}
\put(117.50,31.50){\makebox(0,0)[cc]{1}}
\put(117.50,16.50){\makebox(0,0)[cc]{1}}
\put(117.50,16.52){\circle{3.90}}
\put(122.50,46.50){\makebox(0,0)[cc]{9}}
\put(122.50,41.50){\makebox(0,0)[cc]{7}}
\put(122.50,36.50){\makebox(0,0)[cc]{$*$}}
\put(122.50,31.50){\makebox(0,0)[cc]{$*$}}
\put(122.50,26.50){\makebox(0,0)[cc]{$*$}}
\put(122.50,21.50){\makebox(0,0)[cc]{$*$}}
\put(122.50,16.50){\makebox(0,0)[cc]{10}}
\put(122.50,11.50){\makebox(0,0)[cc]{$*$}}
\put(117.50,6.50){\makebox(0,0)[cc]{6}}
\put(112.50,6.50){\makebox(0,0)[cc]{6}}
\put(107.50,6.50){\makebox(0,0)[cc]{6}}
\put(102.50,6.50){\makebox(0,0)[cc]{3}}
\put(92.50,1.50){\makebox(0,0)[cc]{(f)}}
\end{picture}
}
\caption{Finding antichains}
\label{f3-4}
\end{figure}
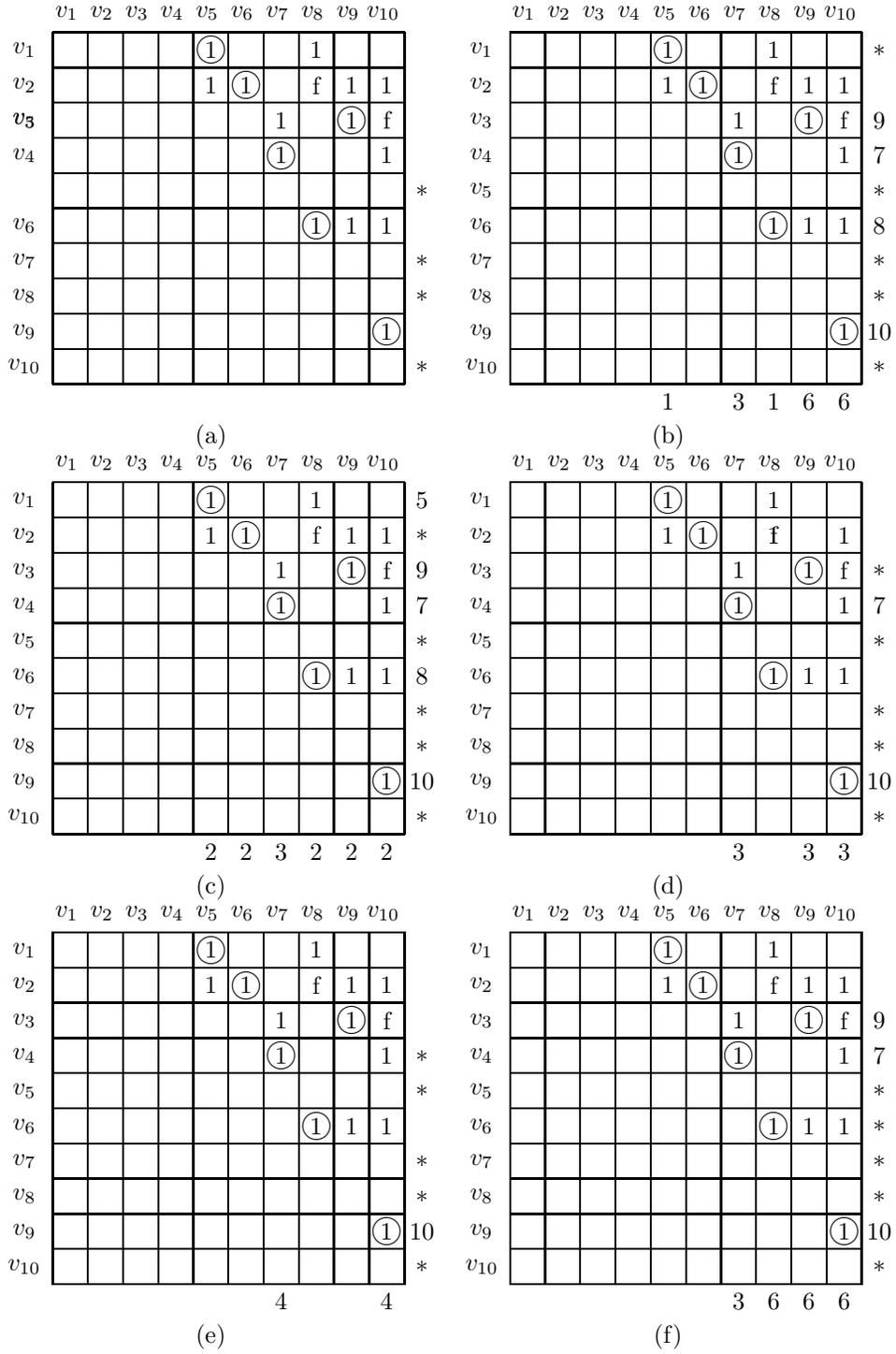

The adjacent matrix of the TCG ${\vec G}_{t}(V^{0})$ is shown in Fig. 
\ref{f3-4}. The fictitious arcs of this graph are represented by the 
letter {\bf f}. Arcs, belonging to the MCP of ${\vec G}_{t}(V^{0})$, 
are put into circles. These arcs are shown in Fig. \ref{f2-4} by thick lines.

First of all, notice that the initiating set $V^{0}$ = 
$\{v_{1},v_{2},v_{3},v_{4}\}$ is a MIS of ${\vec G}(V^{0})$. 

Find the general antichain $\cal U$ (see Fig. \ref{f3-4} (a)). 
The set of marked rows is $U_{r}$ = $\{5,7,8,10\}$, and the set of marked 
columns is empty, that is, $U_{c}= \oslash$. Therefore:
\[
{\cal U}= U_{r}\setminus U_{c}= \{5,7,8,10\}\setminus \oslash =\{5,7,8,10\}.
\]

This antichain is a MIS of the digraph, and $Card({\cal U})$ = $Card(V^{0})$.

Find the maximum antichains ${\cal U}(v)$ for graph vertices. 

Clearly, ${\cal U}(v_{5})$ = ${\cal U}(v_{7})$ = ${\cal U}(v_{8})$ =
${\cal U}(v_{10})$ = $\cal U$.

To find ${\cal U}(v_{1})$, mark the first row of the adjacent matrix 
of ${\vec G}_{t}(X^{0})$.

Marking the first row in Fig. \ref{f3-4} (b), we have $U_{r}$ = 
$\{1, 3, 4, 5, 6, 7, 8. 9. 10\}$, and $U_{c}$ = $\{5, 7, 8, 9, 10\}$.
Consequently, 
\begin{displaymath}
{\cal U}(v_{1}) = U_{r}\setminus U_{c} = \{1, 3, 4, 6\}.
\end{displaymath}

This antichain is a MIS of the digraph, and $Card({\cal U}(v_{1}))$ = 
$Card(X^{0})$.

Similarly, marking the second row in Fig. \ref{f3-4} (c), we have
\begin{displaymath}
{\cal U}(v_{2}) = \{1,2,3,4,5,6,7,8,9,10\}\setminus \{5,6,7,8,9,10\} = 
\{1,2,3,4\}.
\end{displaymath}

This maximum antichain is also a MIS of the digraph, and 
$Card({\cal U}(v_{2}))$ = $Card(V^{0})$.

Similarly, we obtain the maximum antichains 
\[{\cal U}(v_{3}) = \{3, 4, 5, 8\},\]
\[{\cal U}(v_{4}) = \{4, 5, 8, 9\},\] 
and 
\[{\cal U}(v_{6}) = \{3, 4, 5, 6\}\]
from Fig. \ref{f3-4} (d), (e), and (f) correspondingly. Each of these maximum 
antichains is a MIS of the digraph ${\vec G}(V^{0})$, and they have the number 
of elements equal to $V^{0}$.

Thus, the digraph ${\vec G}(V^{0})$ is saturated with respect to
the initiating set $V^{0}$.

Now we examine directed subgraphs induced by layers of ${\vec G}(V^{0})$.

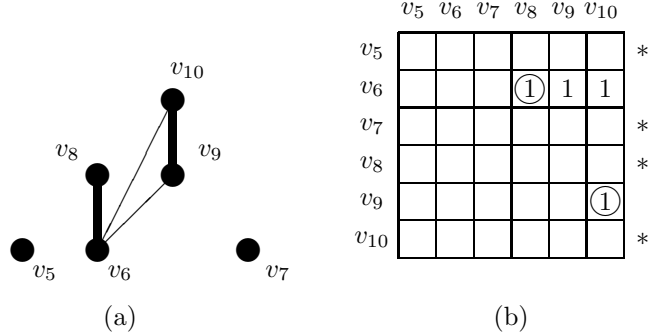
\begin{figure}
\centering
\mbox{\unitlength=1.00mm
\special{em:linewidth 0.4pt}
\linethickness{0.4pt}
\begin{picture}(85.00,43.00)
\put(55.00,10.00){\line(0,1){30.00}}
\put(55.00,40.00){\line(1,0){30.00}}
\put(85.00,40.00){\line(0,-1){30.00}}
\put(85.00,10.00){\line(-1,0){30.00}}
\put(55.00,35.00){\line(1,0){30.00}}
\put(55.00,30.00){\line(1,0){30.00}}
\put(55.00,25.00){\line(1,0){30.00}}
\put(55.00,20.00){\line(1,0){30.00}}
\put(55.00,15.00){\line(1,0){30.00}}
\put(60.00,40.00){\line(0,-1){30.00}}
\put(65.00,40.00){\line(0,-1){30.00}}
\put(70.00,40.00){\line(0,-1){30.00}}
\put(75.00,40.00){\line(0,-1){30.00}}
\put(80.00,40.00){\line(0,-1){30.00}}
\put(51.50,37.50){\makebox(0,0)[cc]{$v_{5}$}}
\put(51.50,32.50){\makebox(0,0)[cc]{$v_{6}$}}
\put(51.50,27.50){\makebox(0,0)[cc]{$v_{7}$}}
\put(51.50,22.50){\makebox(0,0)[cc]{$v_{8}$}}
\put(51.50,17.50){\makebox(0,0)[cc]{$v_{9}$}}
\put(51.50,12.50){\makebox(0,0)[cc]{$v_{10}$}}
\put(57.00,43.00){\makebox(0,0)[cc]{$v_{5}$}}
\put(62.00,43.00){\makebox(0,0)[cc]{$v_{6}$}}
\put(67.00,43.00){\makebox(0,0)[cc]{$v_{7}$}}
\put(72.00,43.00){\makebox(0,0)[cc]{$v_{8}$}}
\put(77.00,43.00){\makebox(0,0)[cc]{$v_{9}$}}
\put(82.00,43.00){\makebox(0,0)[cc]{$v_{10}$}}
\put(82.50,32.50){\makebox(0,0)[cc]{1}}
\put(77.50,32.50){\makebox(0,0)[cc]{1}}
\put(72.50,32.50){\makebox(0,0)[cc]{1}}
\put(72.50,32.52){\circle{3.90}}
\put(82.50,17.50){\makebox(0,0)[cc]{1}}
\put(82.50,17.52){\circle{3.90}}
\put(87.50,37.50){\makebox(0,0)[cc]{$*$}}
\put(87.50,27.50){\makebox(0,0)[cc]{$*$}}
\put(87.50,22.50){\makebox(0,0)[cc]{$*$}}
\put(87.50,12.50){\makebox(0,0)[cc]{$*$}}
\put(5.00,11.00){\circle*{3.00}}
\put(15.00,11.00){\line(1,1){10.00}}
\put(25.00,31.00){\line(-1,-2){10.00}}
\put(15.00,21.00){\circle*{3.00}}
\put(15.00,11.00){\circle*{3.00}}
\put(25.00,21.00){\circle*{3.00}}
\put(25.00,31.00){\circle*{3.00}}
\put(35.00,11.00){\circle*{3.00}}
\put(8.00,8.00){\makebox(0,0)[cc]{$v_{5}$}}
\put(18.00,8.00){\makebox(0,0)[cc]{$v_{6}$}}
\put(39.00,8.00){\makebox(0,0)[cc]{$v_{7}$}}
\put(11.00,24.00){\makebox(0,0)[cc]{$v_{8}$}}
\put(30.00,24.00){\makebox(0,0)[cc]{$v_{9}$}}
\put(27.00,35.00){\makebox(0,0)[cc]{$v_{10}$}}
\put(18.00,2.00){\makebox(0,0)[cc]{(a)}}
\put(70.00,2.00){\makebox(0,0)[cc]{(b)}}
\linethickness{3.0pt}
\put(25.00,21.00){\line(0,1){10.00}}
\put(15.00,11.00){\line(0,1){10.00}}
\end{picture}
}
\caption{The directed subgraph induced by the layer $V^{1}$}
\label{f4-4}
\end{figure}

Consider directed subgraph ${\vec G}(V^{1})$ induced by layer
$V^{1}$ = $\{v_{5},v_{6},v_{7}\}$. This subgraph is represented in 
Fig. \ref{f4-4} (a). Notice that the adjacent matrix of the TCG
${\vec G}_{t}(V^{1})$ can be obtained from the adjacent matrix of
${\vec G}_{t}(V^{0})$ directly. Obviously, the MCP of 
${\vec G}_{t}(V^{1})$ is a part of the MCP of ${\vec G}_{t}(V^{0})$.
The adjacent matrix of ${\vec G}_{t}(V^{1})$ is shown in Fig. \ref{f4-4} (b).
The general antichain ${\cal U}_{1}$ of ${\vec G}_{t}(V^{1})$ equals
\begin{displaymath}
{\cal U}_{1}= \{v_{5},v_{7},v_{8},v_{10}\}.
\end{displaymath}

This antichain is a MIS of ${\vec G}(V^{1})$; however,
$Card({\cal U}_{1})>Card(V^{1})$. Consequently, the directed subgraph
${\vec G}(V^{1})$ is not vertex-saturated with respect to its initiating set.

\begin{figure}
\centering
\mbox{\unitlength=1.00mm
\special{em:linewidth 0.4pt}
\linethickness{0.4pt}
\begin{picture}(85.00,43.00)
\put(55.00,10.00){\line(0,1){30.00}}
\put(55.00,40.00){\line(1,0){30.00}}
\put(85.00,40.00){\line(0,-1){30.00}}
\put(85.00,10.00){\line(-1,0){30.00}}
\put(55.00,35.00){\line(1,0){30.00}}
\put(55.00,30.00){\line(1,0){30.00}}
\put(55.00,25.00){\line(1,0){30.00}}
\put(55.00,20.00){\line(1,0){30.00}}
\put(55.00,15.00){\line(1,0){30.00}}
\put(60.00,40.00){\line(0,-1){30.00}}
\put(65.00,40.00){\line(0,-1){30.00}}
\put(70.00,40.00){\line(0,-1){30.00}}
\put(75.00,40.00){\line(0,-1){30.00}}
\put(80.00,40.00){\line(0,-1){30.00}}
\put(51.50,37.50){\makebox(0,0)[cc]{$v_{5}$}}
\put(51.50,32.50){\makebox(0,0)[cc]{$v_{6}$}}
\put(51.50,27.50){\makebox(0,0)[cc]{$v_{7}$}}
\put(51.50,22.50){\makebox(0,0)[cc]{$v_{8}$}}
\put(51.50,17.50){\makebox(0,0)[cc]{$v_{9}$}}
\put(51.50,12.50){\makebox(0,0)[cc]{$v_{10}$}}
\put(57.00,43.00){\makebox(0,0)[cc]{$v_{5}$}}
\put(62.00,43.00){\makebox(0,0)[cc]{$v_{6}$}}
\put(67.00,43.00){\makebox(0,0)[cc]{$v_{7}$}}
\put(72.00,43.00){\makebox(0,0)[cc]{$v_{8}$}}
\put(77.00,43.00){\makebox(0,0)[cc]{$v_{9}$}}
\put(82.00,43.00){\makebox(0,0)[cc]{$v_{10}$}}
\put(77.50,32.50){\makebox(0,0)[cc]{1}}
\put(77.50,32.52){\circle{3.90}}
\put(62.50,22.50){\makebox(0,0)[cc]{1}}
\put(62.50,22.52){\circle{3.90}}
\put(62.50,12.50){\makebox(0,0)[cc]{1}}
\put(77.50,12.50){\makebox(0,0)[cc]{1}}
\put(77.50,22.50){\makebox(0,0)[cc]{f}}
\put(87.50,37.50){\makebox(0,0)[cc]{$*$}}
\put(87.50,32.50){\makebox(0,0)[cc]{9}}
\put(87.50,27.50){\makebox(0,0)[cc]{$*$}}
\put(87.50,22.50){\makebox(0,0)[cc]{6}}
\put(87.50,17.50){\makebox(0,0)[cc]{$*$}}
\put(87.50,12.50){\makebox(0,0)[cc]{$*$}}
\put(77.50,7.50){\makebox(0,0)[cc]{10}}
\put(62.50,7.50){\makebox(0,0)[cc]{10}}
\put(23.11,2.00){\makebox(0,0)[cc]{(a)}}
\put(69.11,2.00){\makebox(0,0)[cc]{(b)}}
\put(5.00,12.89){\circle*{3.00}}
\put(15.00,32.89){\line(3,-4){15.00}}
\put(30.13,12.89){\line(-3,2){15.00}}
\put(15.00,32.89){\circle*{3.00}}
\put(15.00,22.89){\circle*{3.00}}
\put(15.00,12.89){\circle*{3.00}}
\put(30.00,12.89){\circle*{3.00}}
\put(40.00,12.89){\circle*{3.00}}
\put(8.00,9.89){\makebox(0,0)[cc]{$v_{5}$}}
\put(18.00,9.89){\makebox(0,0)[cc]{$v_{8}$}}
\put(28.00,9.89){\makebox(0,0)[cc]{$v_{10}$}}
\put(43.00,9.89){\makebox(0,0)[cc]{$v_{7}$}}
\put(10.00,24.89){\makebox(0,0)[cc]{$v_{6}$}}
\put(10.00,36.89){\makebox(0,0)[cc]{$v_{9}$}}
\linethickness{3.0pt}
\put(15.00,12.89){\line(0,1){20.00}}
\end{picture}
}
\caption{The new directed subgraph}
\label{f4-5}
\end{figure}
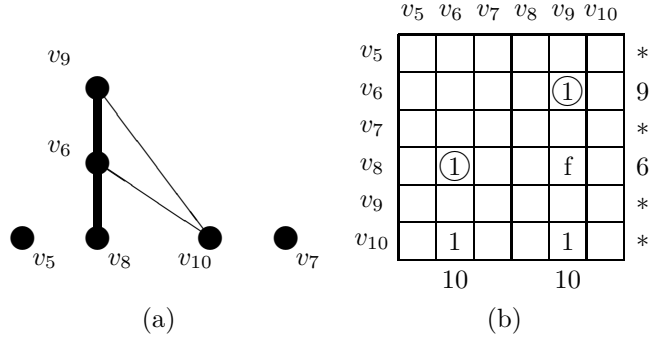

Therefore, we assume $W=\{v_{5}, v_{7}, v_{8}, v_{10}\}$ and
reorientate all arcs of ${\vec G}(X^{1})$ incoming to the vertices of $W$.
We obtain a new directed subgraph represented in Fig. \ref{f4-5} (a). 
Clearly, this subgraph has a new initiating set $V^{1}$ = 
$\{v_{5}, v_{7}, v_{8}, v_{10}\}$.

Examining as above, we find that the new directed subgraph ${\vec G}(V^{1})$ 
is vertex-saturated with respect to its initiating set $V^{1}$.
  
Thus, we may construct a new digraph ${\vec G}(V^{0})$. The adjacent matrix of
this digraph can be obtained from the adjacent matrix of the initial digraph
if the corresponding part of it is replaced by the adjacent matrix
of ${\vec G}(V^{1})$.

Similarly, we determine that a directed subgraph ${\vec G}(V^{2})$, 
where $V^{2}$ = $\{6\}$, is vertex-saturated with respect to
its initiating set $V^{2}$.

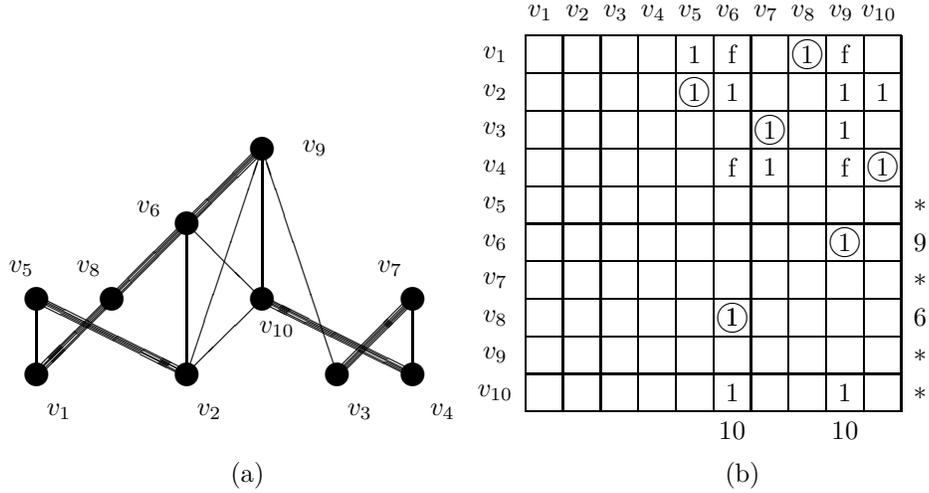
\begin{figure}[hbp]
\centering
\mbox{\unitlength 1.00mm
\linethickness{0.4pt}
\begin{picture}(122.50,63.00)
\put(70.00,10.00){\line(0,1){50.00}}
\put(70.00,60.00){\line(1,0){50.00}}
\put(120.00,60.00){\line(0,-1){50.00}}
\put(120.00,10.00){\line(-1,0){50.00}}
\put(70.00,55.00){\line(1,0){50.00}}
\put(70.00,50.00){\line(1,0){50.00}}
\put(70.00,45.00){\line(1,0){50.00}}
\put(70.00,40.00){\line(1,0){50.00}}
\put(70.00,35.00){\line(1,0){50.00}}
\put(70.00,30.00){\line(1,0){50.00}}
\put(70.00,25.00){\line(1,0){50.00}}
\put(70.00,20.00){\line(1,0){50.00}}
\put(70.00,15.00){\line(1,0){50.00}}
\put(75.00,60.00){\line(0,-1){50.00}}
\put(80.00,60.00){\line(0,-1){50.00}}
\put(85.00,60.00){\line(0,-1){50.00}}
\put(90.00,60.00){\line(0,-1){50.00}}
\put(95.00,60.00){\line(0,-1){50.00}}
\put(100.00,60.00){\line(0,-1){50.00}}
\put(105.00,60.00){\line(0,-1){50.00}}
\put(110.00,60.00){\line(0,-1){50.00}}
\put(115.00,60.00){\line(0,-1){50.00}}
\put(66.00,57.50){\makebox(0,0)[cc]{$v_{1}$}}
\put(66.00,52.50){\makebox(0,0)[cc]{$v_{2}$}}
\put(66.00,47.50){\makebox(0,0)[cc]{$v_{3}$}}
\put(66.00,42.50){\makebox(0,0)[cc]{$v_{4}$}}
\put(66.00,37.50){\makebox(0,0)[cc]{$v_{5}$}}
\put(66.00,32.50){\makebox(0,0)[cc]{$v_{6}$}}
\put(66.00,27.50){\makebox(0,0)[cc]{$v_{7}$}}
\put(66.00,22.50){\makebox(0,0)[cc]{$v_{8}$}}
\put(66.00,17.50){\makebox(0,0)[cc]{$v_{9}$}}
\put(66.00,12.50){\makebox(0,0)[cc]{$v_{10}$}}
\put(72.00,63.00){\makebox(0,0)[cc]{$v_{1}$}}
\put(77.00,63.00){\makebox(0,0)[cc]{$v_{2}$}}
\put(82.00,63.00){\makebox(0,0)[cc]{$v_{3}$}}
\put(87.00,63.00){\makebox(0,0)[cc]{$v_{4}$}}
\put(92.00,63.00){\makebox(0,0)[cc]{$v_{5}$}}
\put(97.00,63.00){\makebox(0,0)[cc]{$v_{6}$}}
\put(102.00,63.00){\makebox(0,0)[cc]{$v_{7}$}}
\put(107.00,63.00){\makebox(0,0)[cc]{$v_{8}$}}
\put(112.00,63.00){\makebox(0,0)[cc]{$v_{9}$}}
\put(117.00,63.00){\makebox(0,0)[cc]{$v_{10}$}}
\put(92.50,57.50){\makebox(0,0)[cc]{1}}
\put(97.50,57.50){\makebox(0,0)[cc]{f}}
\put(107.50,57.50){\makebox(0,0)[cc]{1}}
\put(107.50,57.52){\circle{3.90}}
\put(112.50,57.50){\makebox(0,0)[cc]{f}}
\put(92.50,52.50){\makebox(0,0)[cc]{1}}
\put(92.50,52.52){\circle{3.90}}
\put(97.50,52.50){\makebox(0,0)[cc]{1}}
\put(112.50,52.50){\makebox(0,0)[cc]{1}}
\put(117.50,52.50){\makebox(0,0)[cc]{1}}
\put(102.50,47.52){\circle{3.90}}
\put(102.50,47.50){\makebox(0,0)[cc]{1}}
\put(112.50,47.50){\makebox(0,0)[cc]{1}}
\put(97.50,42.50){\makebox(0,0)[cc]{f}}
\put(102.50,42.50){\makebox(0,0)[cc]{1}}
\put(112.50,42.50){\makebox(0,0)[cc]{f}}
\put(117.50,42.52){\circle{3.90}}
\put(117.50,42.50){\makebox(0,0)[cc]{1}}
\put(112.50,32.50){\makebox(0,0)[cc]{1}}
\put(112.50,32.52){\circle{3.90}}
\put(97.50,22.50){\makebox(0,0)[cc]{1}}
\put(97.50,22.52){\circle{3.90}}
\put(112.50,12.50){\makebox(0,0)[cc]{1}}
\put(97.50,12.50){\makebox(0,0)[cc]{1}}
\put(122.50,37.50){\makebox(0,0)[cc]{$*$}}
\put(122.50,32.50){\makebox(0,0)[cc]{9}}
\put(122.50,27.50){\makebox(0,0)[cc]{$*$}}
\put(97.50,22.50){\makebox(0,0)[cc]{1}}
\put(122.50,22.50){\makebox(0,0)[cc]{6}}
\put(122.50,17.50){\makebox(0,0)[cc]{$*$}}
\put(122.50,12.50){\makebox(0,0)[cc]{$*$}}
\put(112.50,7.50){\makebox(0,0)[cc]{10}}
\put(97.50,7.50){\makebox(0,0)[cc]{10}}
\put(33.06,1.50){\makebox(0,0)[cc]{(a)}}
\put(98.89,1.50){\makebox(0,0)[cc]{(b)}}
\put(5.00,15.00){\line(0,1){10.00}}
\put(5.00,25.00){\line(2,-1){20.00}}
\put(5.15,25.15){\line(2,-1){20.00}}
\put(5.30,25.30){\line(2,-1){20.00}}
\put(4.85,24.85){\line(2,-1){20.00}}
\put(4.70,24.70){\line(2,-1){20.00}}
\put(25.00,15.00){\line(0,1){20.00}}
\put(25.00,35.00){\line(1,-1){10.00}}
\put(35.00,25.00){\line(-1,-1){10.00}}
\put(25.00,15.00){\line(1,3){10.00}}
\put(35.00,45.00){\line(-1,-1){30.00}}
\put(35.15,44.85){\line(-1,-1){30.00}}
\put(35.30,44.70){\line(-1,-1){30.00}}
\put(34.85,45.15){\line(-1,-1){30.00}}
\put(34.70,45.30){\line(-1,-1){30.00}}
\put(45.00,15.00){\line(1,1){10.00}}
\put(45.15,14.85){\line(1,1){10.00}}
\put(45.30,14.70){\line(1,1){10.00}}
\put(44.85,15.15){\line(1,1){10.00}}
\put(44.70,15.30){\line(1,1){10.00}}
\put(35.00,45.00){\line(0,-1){20.00}}
\put(5.00,15.00){\circle*{3.00}}
\put(25.00,15.00){\circle*{3.00}}
\put(45.00,15.00){\circle*{3.00}}
\put(55.00,15.00){\circle*{3.00}}
\put(5.00,25.00){\circle*{3.00}}
\put(15.00,25.00){\circle*{3.00}}
\put(35.00,25.00){\circle*{3.00}}
\put(55.00,25.00){\circle*{3.00}}
\put(25.00,35.00){\circle*{3.00}}
\put(35.00,45.00){\circle*{3.00}}
\put(8.00,10.00){\makebox(0,0)[cc]{$v_{1}$}}
\put(28.00,10.00){\makebox(0,0)[cc]{$v_{2}$}}
\put(48.00,10.00){\makebox(0,0)[cc]{$v_{3}$}}
\put(59.00,10.00){\makebox(0,0)[cc]{$v_{4}$}}
\put(3.00,29.00){\makebox(0,0)[cc]{$v_{5}$}}
\put(12.00,29.00){\makebox(0,0)[cc]{$v_{8}$}}
\put(37.00,21.00){\makebox(0,0)[cc]{$v_{10}$}}
\put(52.00,29.00){\makebox(0,0)[cc]{$v_{7}$}}
\put(20.00,37.00){\makebox(0,0)[cc]{$v_{6}$}}
\put(42.00,45.00){\makebox(0,0)[cc]{$v_{9}$}}
\put(55.00,25.00){\line(0,-1){10.00}}
\put(45.00,15.00){\line(-1,3){10.00}}
\put(55.00,15.00){\line(-2,1){20.00}}
\put(55.15,15.15){\line(-2,1){20.00}}
\put(55.30,15.30){\line(-2,1){20.00}}
\put(54.85,14.85){\line(-2,1){20.00}}
\put(54.70,14.70){\line(-2,1){20.00}}
\end{picture}
}
\caption{The VS-digraph}
\label{f4-6}
\end{figure}

At last, we may make sure that the new digraph ${\vec G}(V^{0})$
is a VS-digraph since each of its directed subgraphs is vertex-saturated with
respect to its initiating set. This digraph is represented in Fig. \ref{f4-6} 
(a). The adjacent matrix of the transitive closure graph ${\vec G}_{t}(V^{0})$ 
and its MCP together are represented in Fig. \ref{f4-6} (b).

\end{document}